\documentclass[%
 reprint,
 superscriptaddress,
nofootinbib,
 amsmath,amssymb,
 aps, physrev,
]{revtex4-2}

\usepackage{graphicx}
\usepackage{dcolumn}
\usepackage{bm}
\usepackage{hyperref}
\usepackage{mathrsfs}

\usepackage{xcolor}

\begin{document}


\title{\textbf{Linear Analysis and Simulations of the Cosmic-Ray Streaming Instability: the Importance of Oblique Waves} 
}%

\author{Shuzhe Zeng}
\email{szeng23@umd.edu}
\affiliation{
 Department of Physics, Tsinghua University, Beijing 100084, China
}%
\affiliation{
 Department of Physics, University of Maryland, College Park, MD 20742, USA
}%
\affiliation{
 Institute for Research in Electronics and Applied Physics, University of Maryland, College Park, MD 20742, USA
}%
\author{Xue-Ning Bai}
\email{xbai@tsinghua.edu.cn}
\affiliation{
 Institute for Advanced Study, Tsinghua University, Beijing 100084, China
}%
\affiliation{
Department of Astronomy, Tsinghua University, Beijing 100084, China
}%
\author{Xiaochen Sun}
\affiliation{
 Institute for Advanced Study, Tsinghua University, Beijing 100084, China
}%
\affiliation{
 Department of Astrophysical Sciences, Princeton University, Princeton, NJ 08544, USA
}%

\date{\today}

\begin{abstract}
Cosmic-ray (CR) streaming instability (CRSI) is believed to play an important role in CR transport and CR feedback to galaxies. It drives the growth of magnetohydrodynamic (MHD) waves that scatter CRs, and leads to energy/momentum exchange between CRs and interstellar medium. Despite extensive research on CRSI, its dependence on the thermodynamic state of the gas and its multidimensional effects have not been systematically studied. In this study, we derive the dispersion relation of the CRSI for three types of MHD waves including their dependence on propagation direction and plasma $\beta$ (the ratio of thermal pressure to magnetic pressure). We verify the analytical dispersion relation with one-dimensional and two-dimensional magnetohydrodynamic particle-in-cell simulations. Furthermore, we use 2D simulations to investigate the role of oblique MHD waves in scattering CRs, and find that these waves are important in helping low-energy particles overcome the 90-degree pitch angle barrier.
While magnetosonic waves tend to be damped by transit time damping under typical conditions, oblique Alfv\'en waves likely play an important role in low-$\beta$ plasmas.
\end{abstract}

\maketitle


\section{Introduction}

Cosmic rays (CRs) are (trans-)relativistic high-energy charged particles. In the Galaxy, the dominant CR population is $\sim$GeV protons with a number density of $n_{\mathrm{CR}} \sim 1 \times 10^{-9} \mathrm{cm}^{-3}$ \citep{Zweibel2017}. For comparison, the number density of background ions in the interstellar medium (ISM) is $n_i \sim 1\mathrm{cm}^{-3}$. Despite the low number density, the energy density of CRs is significant, measuring around $1\mathrm{eV\, cm}^{-3}$, which is comparable to the energy density of gases, magnetic fields, and turbulence in the ISM \citep{Ferriere2001}. Therefore, CRs can play an important role in the dynamical processes of galaxy formation and evolution (known as CR feedback). In addition to providing pressure support against gravity, CRs are well known to be able to drive galactic winds \cite[e.g.][]{Ipavich_1975, Breitschwerdt_1991, Zirakashvili_1996, Hopkins2020, Quataert2022}, and also serve as a crucial source of ionization and heating in dense regions of the ISM \citep{Grenier2015, Zweibel2017}.

Measurements of CR composition reveal that $\sim$GeV CRs reside in the Galaxy for $1\text{--}2\times 10^7$ years, which is much longer than the light crossing time of the Galaxy, implying that CRs are efficiently confined \citep{Zweibel_2013}. The confinement is attributed to scattering on magnetic fluctuations. The scattering of high-energy CRs \cite[above a few hundred GeV;][]{Blasi_2012} is likely dominated by external turbulence \cite[e.g.][]{Jokipii_1966, Schlickeiser_1998, yan_2002}. For $\sim$GeV CRs, these fluctuations are widely believed to be self-generated magnetohydrodynamic (MHD) waves. The physical mechanism underlying this process is the gyroresonant cosmic-ray streaming instability (CRSI).
When the bulk drift speed of the CRs exceeds the phase velocity of MHD waves, the free energy stored in CR streaming can be transferred to the background gas through resonant interactions between the gyrating CRs and the MHD waves, thus driving their growth \citep{Kulsrud1969, Wentzel_1974, Skilling1, Skilling2, Skilling3}.\footnote{We note that in the limit of extremely strong CR streaming, the streaming CRs leads to a strong CR current that substantially enhances the right-handed Alfv\'en mode, whose dominance is known as the Bell instability \citep{Bell_2004} and is non-resonant (current-driven) in nature. In this work, we do not consider this regime.}
By enabling energy and momentum exchange between CRs and the ISM gas,the CRSI is believed to play a crucial role in CR feedback (e.g., Ruszkowski \& Pfrommer 2023).

The CRSI has been mostly studied as a 1D instability which excites parallel-propagating Alfv\'en waves of both polarizations along the direction of CR streaming \cite[e.g.][]{Achterberg1981, BLANDFORD_1987}. The growth rates for parallel-propagating modes are fully characterized by the CR distribution function and the Alfv\'en speed. However, the role of oblique modes have not been systematically investigated. These modes not only include the Alfv\'en waves, but also the fast and slow magnetosonic waves, which can also resonantly interact with the CR particles under certain conditions. The interaction of the CRs with oblique MHD waves would enable additional wave-particle coupling, potentially alter our understandings of CR feedback. Moreover, the incorporation of the fast and slow modes introduces another parameter, the sound speed, into the problem, which can be characterized by plasma $\beta$ (ratio of gas to magnetic pressure).

Early studies by \cite{Kulsrud1969} and \cite{Tademaru_1969} already showed the full dispersion relation of waves in the general propagation direction, indicating that the CRSI can be generated for both Alfv\'en and fast magnetosonic waves over a broad range of wave obliquities. 
The resonant scattering of particles by a given distribution of waves, including oblique ones, was examined in \cite{Melrose_1969}. However, these analyses were restricted to the cold plasma limit. While \cite{Melrose_1975} incorporated finite sound speed and discussed all MHD modes, the focus was primarily on the slow mode in a low-$\beta$ condition. 
On the other hand, \cite{Achterberg1981} considered the high-$\beta$ limit, but only focused on waves propagating parallel to the background magnetic field. 
The lack of theoretical investigation of oblique modes in a general plasma-$\beta$ condition is particularly limiting given that the plasmas in the ISM and circumgalactic medium (CGM) are multiphase in nature, encompassing a wide range of physical conditions (e.g., $\beta$ values) across and beyond the Galaxy \citep{Ferriere2001, Oh2023}. Therefore, the oblique CRSI could potentially operate in different ways in different environments. In this work, our first goal is to derive and systematically study the linear dispersion relation of the CRSI in the most general case (general $\beta$ and oblique propagation), filling in a major gap in the literature.

In recent years, it has become possible to directly simulate the CRSI to not only verify the linear theory but also study the subsequent evolution of both particles and waves \citep{Bai2019,Holcomb_2019}. The problem is difficult for standard particle-in-cell (PIC) approaches, as the requirement to resolve the background plasma scale leaves substantial scale separation to accommodate the CRs, and simulations typically adopt relatively extreme parameters with $\delta$-function-type CR distribution functions \citep{Holcomb_2019, Shalaby_2021}. By treating electrons as fluid, the hybrid code partially alleviates the scale separation problem \cite[e.g.][]{Haggerty2019, Benedikt2025}. The issue of scale separation is largely overcome thanks to the use of magnetohydrodynamic particle-in-cell (MHD-PIC) method \citep{Bai_2015}: by treating the background plasma as a fluid described by MHD, one can better focus on the dynamics at CR-gyro-radius scale. When supplemented by the $\delta f$ weighting scheme, one can further suppress Poisson noise, allowing for the accurate reproduction of CRSI growth rates over a broad range of wavelengths \citep{Bai2019}. This opens the avenue to incorporate more physics \cite[e.g.][]{Plotnikov_2021} and to faithfully follow its subsequent evolution \cite[e.g.][]{Bai_2022, Bambic_2021}.

In this paper, following the derivation of the most general dispersion relation of CRSI, we further verify the analytical results using one-dimensional and two-dimensional MHD-PIC simulations. In particular, our 2D simulations are the first of this type, allowing us to follow the subsequent quasi-linear evolution of CR particles with a full spectrum of waves. While CRs are expected to get isotropized over time, there is the well-known 90-degree pitch angle barrier where quasi-linear theory (QLT) fails (as resonant wavelength goes to zero). The presence of oblique waves leads to the formation of weakly oblique discontinuity structures that we find to play an important role in helping CR particles cross the 90-degree barrier, especially for low-energy CRs.

This paper is organized as follows. In Section~\ref{sec: derivation}, we derive the dispersion relation for CRSI under a general plasma $\beta$. A more in-depth analysis of the underlying physics is presented in Section~\ref{sec:phy analy}. We describe the setup of 1D and 2D simulations in Section~\ref{sec:simulation setup} and present the simulation results in Section~\ref{sec:sim results}. We conclude with further discussions in Section~\ref{sec:dis and conc}. Additional derivations and explanations are provided in Appendix~\ref{appen A},~\ref{appen B} and~\ref{appen C}.

\section{Formulation and Derivations}
\label{sec: derivation}

In this section, we present the general form of the growth rates of three MHD wave modes with basic derivations. The results are obtained in the $n_{\mathrm{CR}}/n_i \ll 1$ limit (see Appendix~\ref{appen A} for a more general result). We assume that the CRs and the background ions have the same composition in this paper, but this can be generalized easily. In the $n_{\mathrm{CR}}/n_i \ll 1$ limit, eigen modes of the system are ordinary MHD waves with slight modifications: There is an additional imaginary part in the frequency of the waves,
\begin{equation}
    \omega(k) = \omega_0(k) + \mathrm{i} \Gamma(k),
\end{equation}
where $\omega_0$ is the ordinary frequency without CRs and $\Gamma$ is the additional imaginary part. Since $\Gamma \ll \omega_0$, $\omega_0 \approx \omega$ and we will no longer distinguish between them. The emergence of $\Gamma$ means that the waves would grow or damp.

\subsection{MHD Wave Modes: Basic Geometry}
\begin{figure}
    \includegraphics[width=\columnwidth]{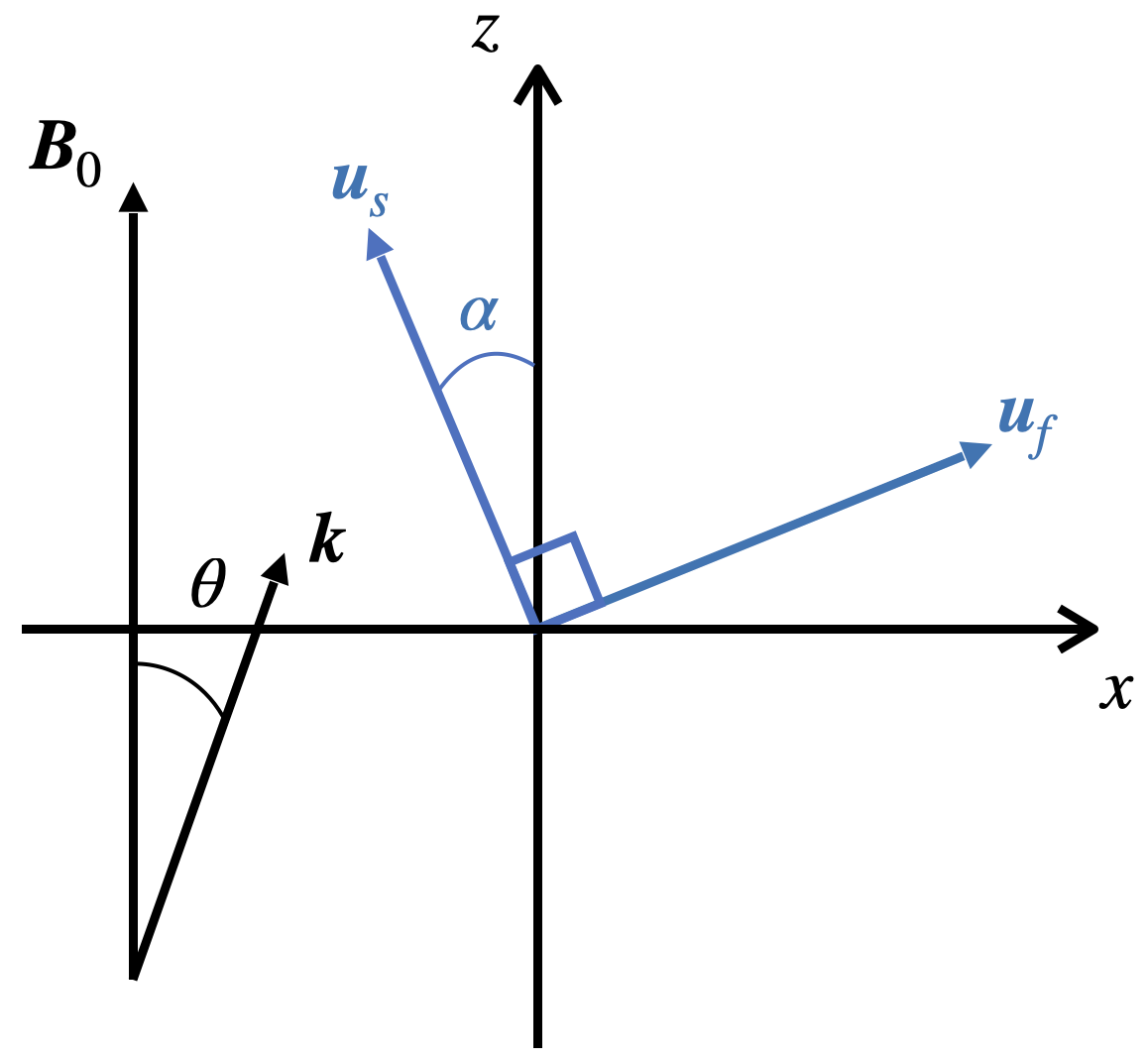}
    \caption{Geometry of the system. The background field and wave vector are denoted by $\boldsymbol{B_0}$ and $\boldsymbol{k}$, with $\theta$ being the angle between them. Both of them lie in the x-z plane and $\boldsymbol{B_0}$ is along $z$ axis. The perturbed velocities of fast and slow mode are represented by $\boldsymbol{u}_f$ and $\boldsymbol{u}_s$, and they are perpendicular to each other. The angle between $\boldsymbol{u}_s$ and $\boldsymbol{B}_0$ is denoted by $\alpha$, which is related to the ratio of the electric field energy to the total wave energy. The perturbed velocity of the Alfven mode is along $\Hat{y}$, which is perpendicular to the plane shown in the figure.}
    \label{fig:geometry}
\end{figure}
Consider a static background gas with constant density $\rho_0$ and magnetic field $\boldsymbol{B}_0 = B_0 \Hat{z}$ with $B_0 >0$. It is well-known that in such a system there are three MHD wave modes: Alfven, fast and slow mode (the latter two are magnetosonic modes). For simplicity, consider monochromatic waves in the form of 
\begin{equation}
    q = \tilde{q} \exp[\mathrm{i}(\boldsymbol{k} \cdot \boldsymbol{r} - \omega t)],
\end{equation}
where $q$ represents a perturbed quantity, which could be the perturbed magnetic field $\delta \boldsymbol{B}\equiv \boldsymbol{B} - \boldsymbol{B}_0$ or fluid velocity $\boldsymbol{u}$; $\tilde{q}$ is its amplitude, a complex number that also contains phase information; $\boldsymbol{k}$ and $\omega$ are the wave vector and frequency, respectively. The dispersion relation of these three modes reads,
\begin{equation}
    \omega_i^2 = \left\{
    \begin{array}{c}
    k_\parallel^2 v_A^2\\[6pt]
    k^2 \, \big[v_A^2 + c_s^2 + \sqrt{(v_A^2 + c_s^2)^2 - 4v_A^2 c_s^2 \cos^2\theta} \big]/2\\[6pt]
    k^2 \, \big[v_A^2 + c_s^2 - \sqrt{(v_A^2 + c_s^2)^2 - 4v_A^2 c_s^2 \cos^2\theta} \big]/2
    \end{array} \right\},
    \label{eq:MHD wave}
\end{equation}
where $\theta$ refers to the angle between background field $\boldsymbol{B}_0$ and wave vector $\boldsymbol{k}$, $v_A = B_0/\sqrt{4\pi\rho_0}$ is the Alfven speed and $c_s$ is the sound speed. The top, middle and bottom lines correspond to the Alfven mode ($i=a$), fast mode ($i=f$) and slow mode ($i=s$), respectively, and this ordering is maintained throughout this paper.

Without loss of generality, let the wave vector be in the $x\text{--} z$ plane. The geometry of the modes is shown in Figure~\ref{fig:geometry}.
The perturbed velocity of the Alfven mode is along $\Hat{y}$, which is perpendicular to the plane shown in the figure. The perturbed velocities of the fast and slow modes, represented by $\boldsymbol{u}_f$ and $\boldsymbol{u}_s$, are in the $x\text{--} z$ plane, and they are perpendicular to each other. Using the relation $\boldsymbol{E} = -\boldsymbol{u} \times \boldsymbol{B_0}/c$, where $c$ is the speed of light, we see that electric field of the Alfven mode is parallel to the $x$ axis while those of magnetosonic mode are parallel to the $y$ axis. Thus, we can decompose the perturbed electric fields $\boldsymbol{E}$ into the form
\begin{equation}
    \boldsymbol{E} = E_x \,\Hat{x} + E_y \,\Hat{y} = E_a \,\Hat{x} + (E_f + E_s) \,\Hat{y},
\end{equation}
where $E_x$ and $E_y$ represent the components of electric fields along $\Hat{x}$ and $\Hat{y}$. The subscripts "a", "f", and "s" correspond to the Alfven wave, fast wave, and slow wave, respectively.

The angle between $\boldsymbol{u}_s$ and $\boldsymbol{B}_0$ is denoted by $\alpha$. For an Alfven wave, fast wave, or slow wave, the proportion of electric field energy in the total wave energy is $(v_A/c)^2/2, \, \cos^2\alpha\,(v_A/c)^2/2$, or $\sin^2\alpha\,(v_A/c)^2/2$, respectively. One can show that $\alpha$ is determined solely by $\beta$ and $\theta$, and the explicit expression of $\alpha$ is 
\begin{align}
    \cos^2\alpha &= \frac{1}{2} \bigg(1 + \frac{1 - \beta \cos(2\theta)/2}{\sqrt{(1+\beta/2)^2 - 2\beta \cos^2 \theta}}\bigg), \notag \\
    \sin^2\alpha &= \frac{1}{2} \bigg(1 - \frac{1 - \beta \cos(2\theta)/2}{\sqrt{(1+\beta/2)^2 - 2\beta \cos^2 \theta}}\bigg).
    \label{eq:cos_sin_alpha}
\end{align}

\subsection{Interaction between CRs and MHD Waves}
\label{sec:interaction}
When introducing the CRs, the linearized momentum equation of the gas becomes
\begin{align}
        \rho_0 \frac{\partial \boldsymbol{u}}{\partial t} &= \frac{1}{4 \pi}(\nabla \times \delta\boldsymbol{B})\times \boldsymbol{B_0} - \frac{1}{c} \boldsymbol{j}_{\mathrm{CR},0} \times \delta\boldsymbol{B} \notag \\ 
        & - \frac{1}{c} \boldsymbol{j}_{\mathrm{CR},1}\times \boldsymbol{B_0} - e n_{\mathrm{CR}} \boldsymbol{E} - \nabla P,
\end{align}
while the continuity equation, the induction equation and the equation of state remain unchanged. In the above, $e$ is the elementary charge; $\boldsymbol{j}_{\mathrm{CR},0}$ is the current from CR bulk motion and the response of CRs to MHD waves is represented by perturbed CR current, $\boldsymbol{j}_{\mathrm{CR},1}$; $P$ is the thermal pressure. 
In linear analysis, $\boldsymbol{j}_{\mathrm{CR},1}$ is linearly related to electric fields,
\begin{equation}
    \boldsymbol{j}_{\mathrm{CR},1} = -\frac{\mathrm{i} \omega}{4\pi} \boldsymbol{\chi}^{\mathrm{CR}} \cdot \boldsymbol{E},
\end{equation}
where $\boldsymbol{\chi}^{\mathrm{CR}}$ is the CR susceptibility tensor. Growth or damping of the waves is due to the energy exchange between background gas and CRs. The power of the work done by the background gas on the CRs is 
\begin{align}
    \mathscr{P} &= \langle \boldsymbol{j}_{\mathrm{CR},1} \cdot \boldsymbol{E} \rangle \notag \\
    &= -\frac{\omega}{4\pi} \langle \,
    (\mathrm{i} \chi_{xx}^{\mathrm{CR}} E_x + \mathrm{i}\chi_{xy}^{\mathrm{CR}} E_y) \, E_x + (\mathrm{i} \chi_{yx}^{\mathrm{CR}} E_x + \mathrm{i}\chi_{yy}^{\mathrm{CR}} E_y) \, E_y \, \rangle \notag \\
    &= \frac{\omega}{8\pi} \left[ \chi^{\mathrm{CR}}_{xx, I} \, \tilde{E_a}^* \tilde{E_a} + \chi^{\mathrm{CR}}_{yy, I} \, (\tilde{E_f}^* \tilde{E_f} + \tilde{E_s}^* \tilde{E_s})
    \right],
    \label{eq:total power}
\end{align}
where $\langle \dots \rangle$ represents averaging on time and space. The $xx$ and $yy$ components of $\boldsymbol{\chi}^{\mathrm{CR}}$ are denoted by $\chi^{\mathrm{CR}}_{xx}$ and $\chi^{\mathrm{CR}}_{yy}$, and $\chi^{\mathrm{CR}}_{xx, I}$, $\chi^{\mathrm{CR}}_{yy, I}$ are their imaginary parts. In the above, we have used the formula
\begin{equation}
    \langle q_1 q_2\rangle = \frac{1}{4}(\tilde{q_1}^* \tilde{q_2} + \tilde{q_1} \tilde{q_2}^*) \; \delta(\boldsymbol{k_1}, \boldsymbol{k_2}) \; \delta(\omega_1, \omega_2).
\end{equation}
The two $\delta$-functions imply that these waves evolve independently: the cross terms vanish since these waves have different dispersion relations. Contribution of each mode to the total power is clearly
\begin{equation}
    \mathscr{P}_i = \frac{\omega}{8\pi}\left\{
    \begin{array}{c}
    \chi^{\mathrm{CR}}_{xx, I} \, \tilde{E_a}^* \tilde{E_a}, \quad i=a \\[6pt]
    \chi^{\mathrm{CR}}_{yy, I} \, \tilde{E_i}^* \tilde{E_i}, \quad i=f, s
    \end{array} \right\}.
    \label{eq:power single mode}
\end{equation}
The total energy density of the wave is 
\begin{equation}
    \mathscr{E} = 2 \, \mathscr{E}_k = \frac{1}{2} \, \rho_0 \, |\tilde{u}|^2,
\end{equation}
where $\mathscr{E}_k = \rho_0 |\tilde{u}|^2/4$ is the density of kinetic energy. For an MHD wave, its total energy is twice its kinetic energy. Based on energy conservation, 
\begin{equation}
    \frac{\mathrm{d} \mathscr{E}}{\mathrm{d} t} = 2 \, \Gamma \, \mathscr{E} = -\mathscr{P}.
\end{equation}
Using Equation~(\ref{eq:power single mode}), we arrive at
\begin{equation}
    \Gamma_i = -\frac{\mathscr{P}_i}{2\mathscr{E}_i} = \left\{
    \begin{array}{c}
    -\big[\dfrac{1}{2} (\dfrac{v_A}{c})^2\big]\,  \chi^{\mathrm{CR}}_{xx, I} \, \omega\\[6pt]
    -\big[\dfrac{1}{2} \cos^2\alpha (\dfrac{v_A}{c})^2 \big]\, \chi^{\mathrm{CR}}_{yy, I} \, \omega\\[6pt]
    -\big[\dfrac{1}{2} \sin^2\alpha (\dfrac{v_A}{c})^2 \big]\, \chi^{\mathrm{CR}}_{yy, I} \, \omega
    \end{array} \right\}.
    \label{eq:growth rate}
\end{equation}
Here $\Gamma$ can be divided into two parts. The first part, $(v_A/c)^2/2, \, \cos^2\alpha\,(v_A/c)^2/2$, and $\sin^2\alpha\,(v_A/c)^2/2$, are the proportions of the electric field energy in the total wave energy of the three modes. Note that this part is unrelated to the presence of the CRs. As mentioned before, the response of CRs to MHD waves is included in $\boldsymbol{\chi}^{\mathrm{CR}}$, the second part of $\Gamma$.

In order to obtain the CR dielectric tensor, we solve the Vlasov equation,
\begin{equation}
    \frac{\partial f}{\partial t} + \boldsymbol{v} \cdot \nabla f + e(\boldsymbol{E} + \frac{\boldsymbol{v} \times \boldsymbol{B}}{c}) \cdot \nabla_p f = 0,
\end{equation}
where $f(\boldsymbol{r}, \boldsymbol{p}, t)$ is the distribution function of CRs, $\boldsymbol{r}$ is the position, $\boldsymbol{p}$ is the momentum, $\boldsymbol{v}$ is the velocities of CRs. Let $f_0(\boldsymbol{p}) = f_0(p_\parallel, p_\perp)$ be the background CR distribution function and $f_1 = f_1(t, \boldsymbol{r}, p_\parallel, p_\perp, \phi)$ be the CR response to the waves, where subscripts $\parallel$ and $\perp$ denote components parallel and perpendicular to background magnetic field, and $\phi$ is the gyro-phase. Thus, $f = f_0 + f_1$. For perturbations of the form $\exp [\mathrm{i} (\boldsymbol{k} \cdot \boldsymbol{r} - \omega t)]$, $f_1$'s dependence on $\boldsymbol{r}$ and $t$ should also be in this form. After some algebra, we obtain the explicit expression of $f_1$ (see \cite{Lifshitz1981} or \cite{Stix1992} for detailed derivation),
\begin{align}
    f_1 = -eA \mathrm{e}^{\mathrm{i} (\boldsymbol{k} \cdot \boldsymbol{r} - \omega t)} &\int_0^{+\infty} [E_x \cos (\phi + \Omega \tau) + E_y \sin (\phi + \Omega \tau)] \notag \\
    & \cdot \mathrm{e}^{\mathrm{i} \left[ (\omega - k_\parallel v_\parallel) \tau - k_\perp v_\perp [\sin (\phi + \Omega \tau) - \sin \phi]/ \Omega \right]}\, \mathrm{d} \tau ,
\end{align}
where $\Omega \equiv eB_0/(\gamma m c)$ is the CR gyro-frequency, $\gamma$ is the Lorentz factor, $m$ is the mass of a CR, and 
\begin{equation}
    A \equiv (1 - \frac{k_\parallel v_\parallel}{\omega}) \frac{\partial f_0}{\partial p_\perp} + \frac{v_\perp k_\parallel}{\omega} \frac{\partial f_0}{\partial p_\parallel}.
\end{equation}
Having obtained $f_1(\boldsymbol{r}, \boldsymbol{p}, t)$, we can find $\boldsymbol{j}_{\mathrm{CR}, 1}$ through
\begin{equation}
    \boldsymbol{j}_{\mathrm{CR},1} = e \int \mathrm{d}^3 \boldsymbol{p} \: \boldsymbol{v} f_1(\boldsymbol{r}, \boldsymbol{p}, t).
\end{equation}
The integral can be done by using the relation 
\begin{align}    
    \int_0^{2\pi} \mathrm{d}\phi \, &\mathrm{e}^{- \mathrm{i} \zeta [\sin (\phi + \Omega \tau) - \sin \phi]} 
    \begin{bmatrix}
    \cos \phi \cos (\phi + \Omega \tau) \\[8pt]
    \sin \phi \sin (\phi + \Omega \tau)
    \end{bmatrix} \notag \\
    &= 2\pi \sum_{n = -\infty}^{+\infty} \mathrm{e}^{-\mathrm{i} n \Omega \tau} 
    \begin{bmatrix}
    n^2 J_n^2(\zeta)/\zeta^2 \\[8pt]
    (J_n'(\zeta))^2
    \end{bmatrix},
\end{align}
where $\zeta \equiv k_\perp v_\perp/\Omega$, $J_n$ is the Bessel function of the first kind of order $n$, and $J_n'$ is the derivative of $J_n$. Then, the dielectric tensor is found to be
\begin{equation}
    \left\{ 
    \begin{array}{c}
         \chi^{\mathrm{CR}}_{xx}  \\
         \chi^{\mathrm{CR}}_{yy}
    \end{array}
    \right\}
     = \frac{4\pi e^2}{\omega} \sum_{n = -\infty}^{+ \infty} \int \mathrm{d}^3 p \, \frac{v_\perp A}{\omega - k_\parallel v_\parallel - n \Omega} 
     \left\{ 
    \begin{array}{c}
         n^2 J_n^2(\zeta)/\zeta^2  \\
         (J_n'(\zeta))^2
    \end{array}
    \right\}.
\end{equation}
Inserting this into Equation~(\ref{eq:growth rate}), we obtain
\begin{align}
    \Gamma_i = & 2 \pi^2 e^2 (\frac{v_A}{c})^2 \sum_{n = -\infty}^{+ \infty} \notag \\
    & \cdot \int \mathrm{d}^3 p \, v_\perp A \, \delta(\omega - k_\parallel v_\parallel - n \Omega) 
    \left\{ 
    \begin{array}{c}
        n^2 J_n^2(\zeta)/\zeta^2 \\
        (J_n'(\zeta))^2 \, \cos^2\alpha \\
        (J_n'(\zeta))^2 \, \sin^2\alpha
    \end{array}
    \right\}.
    \label{eq:grow rate final}
\end{align}
In the above, we have adopted the Landau contour, and used the Plemelj formula:
\begin{equation}
    \lim_{\epsilon \to 0^+} \int \frac{g(z)}{z - \mathrm{i}\epsilon}\, \mathrm{d}z = \mathrm{P} \int \frac{g(z)}{z}\, \mathrm{d}z + \mathrm{i} \pi \int g(z) \delta(z)\, \mathrm{d}z,
\end{equation}
where $g(z)$ is an analytic function, and $\mathrm{P}$ denotes the Cauchy principle value.

\subsection{The Drifting Model}
If the background CR momentum distribution is isotropic, denoted by $F(p)$, in a frame drifting relative to the background gas along the background magnetic field $\boldsymbol{B_0}$ at velocity $v_d \ll c$ (assuming $v_d > 0$ without loss of generality), then $A$ will reduce to (one can find the detailed derivation in Chapter 12 of \cite{Kulsrud2005})
\begin{equation}
    A = \frac{\mathrm{d}F}{\mathrm{d}p} \frac{p_\perp}{p} (1 - \frac{k_\parallel v_d}{\omega}).
    \label{eq:A}
\end{equation}
This is the case we focus on below.

Although the CRSI is commonly analyzed with $F$ being a truncated power law distribution, we adopt a $\kappa$ distribution as $F$ in this paper for two reasons. We have found that the artificial discontinuity of a truncated power law distribution could lead to spurious growth rates when $\theta$ is large (see Appendix ~\ref{appen B} for a detailed discussion). Also, the $\delta f$ method employed by the simulation requires $F$ to be finite at all $p$. The general form of a $\kappa$ distribution reads
\begin{equation}
    F(p) = \frac{n_{\mathrm{CR}}}{(\pi \kappa p_0^2)^{3/2}} \frac{\Gamma(\kappa + 1)}{\Gamma(\kappa - \frac{1}{2})} \bigg[1 + \frac{1}{\kappa} (\frac{p}{p_0})^2 \bigg]^{-(\kappa + 1)}.
    \label{eq:kappa}
\end{equation}
Inserting this and Equation~(\ref{eq:A}) into Equation~(\ref{eq:grow rate final}), we obtain
\begin{align}
    \frac{\Gamma_i}{\Omega_c} &= \; (\frac{k_\parallel v_d}{\omega_i} - 1) \frac{n_\mathrm{CR}}{n_i} \notag \\ & \cdot \frac{2\sqrt{\pi}}{\kappa^{3/2}} \frac{\kappa + 1}{\kappa} \frac{\Gamma(\kappa + 1)}{\Gamma(\kappa - \frac{1}{2})} \sum_{n = -\infty}^{+ \infty} \frac{1}{|\tilde{k}_\parallel|} \int_{|n| \tilde{p}_{\mathrm{res}}}^{+\infty} \mathrm{d} \tilde{p} \, \tilde{p} \bigg[1 + \dfrac{\tilde{p}^2}{\kappa} \bigg]^{-(\kappa + 2)} \notag \\
    & \cdot \left\{ 
    \begin{array}{c}
        \dfrac{n^2}{\tilde{k}_\perp^2} J_n^2(\tilde{k}_\perp \sqrt{\tilde{p}^2 - n^2 \tilde{p}^2_{\mathrm{res}}})\\ [8pt]
        (\tilde{p}^2 - n^2 \tilde{p}_{\mathrm{res}}^2) \bigg[ J_n'(\tilde{k}_\perp \sqrt{\tilde{p}^2 - n^2 \tilde{p}^2_{\mathrm{res}}})\bigg]^2 \, \cos^2\alpha\\
        (\tilde{p}^2 - n^2 \tilde{p}_{\mathrm{res}}^2) \bigg[ J_n'(\tilde{k}_\perp \sqrt{\tilde{p}^2 - n^2 \tilde{p}^2_{\mathrm{res}}})\bigg]^2 \, \sin^2\alpha
    \end{array}
    \right\},
    \label{eq:final}
\end{align}
where $\Omega_c \equiv eB_0/mc$ is the cyclotron frequency; $\tilde{p} \equiv p/p_0, \, \tilde{k}_\parallel \equiv k_\parallel p_0/(m \Omega_c), \, \tilde{k}_\perp \equiv k_\perp p_0/(m \Omega_c), \, \tilde{p}_{\mathrm{res}} \equiv m \Omega_c/(|k_\parallel| \, p_0)$ are dimensionless parameters.

In Equation~(\ref{eq:final}), we see that the sign of the growth rate is determined by the factor $(k_\parallel v_d/\omega - 1)$. The physical meaning of this factor can be referred to in \cite{Bai2019} and \cite{Kulsrud2005}. When $k_\parallel < 0$ (waves propagating backward relative to CR streaming), we always have $\Gamma < 0$, which means wave damping. When $k_\parallel > 0$, $\Gamma > 0$ if $v_d$ exceeds the wave phase speed. In this case, waves can grow, and the following discussion will mainly focus on this scenario.

\section{Physical Analysis}
\label{sec:phy analy}

\begin{figure*}
    \includegraphics[width=17cm]{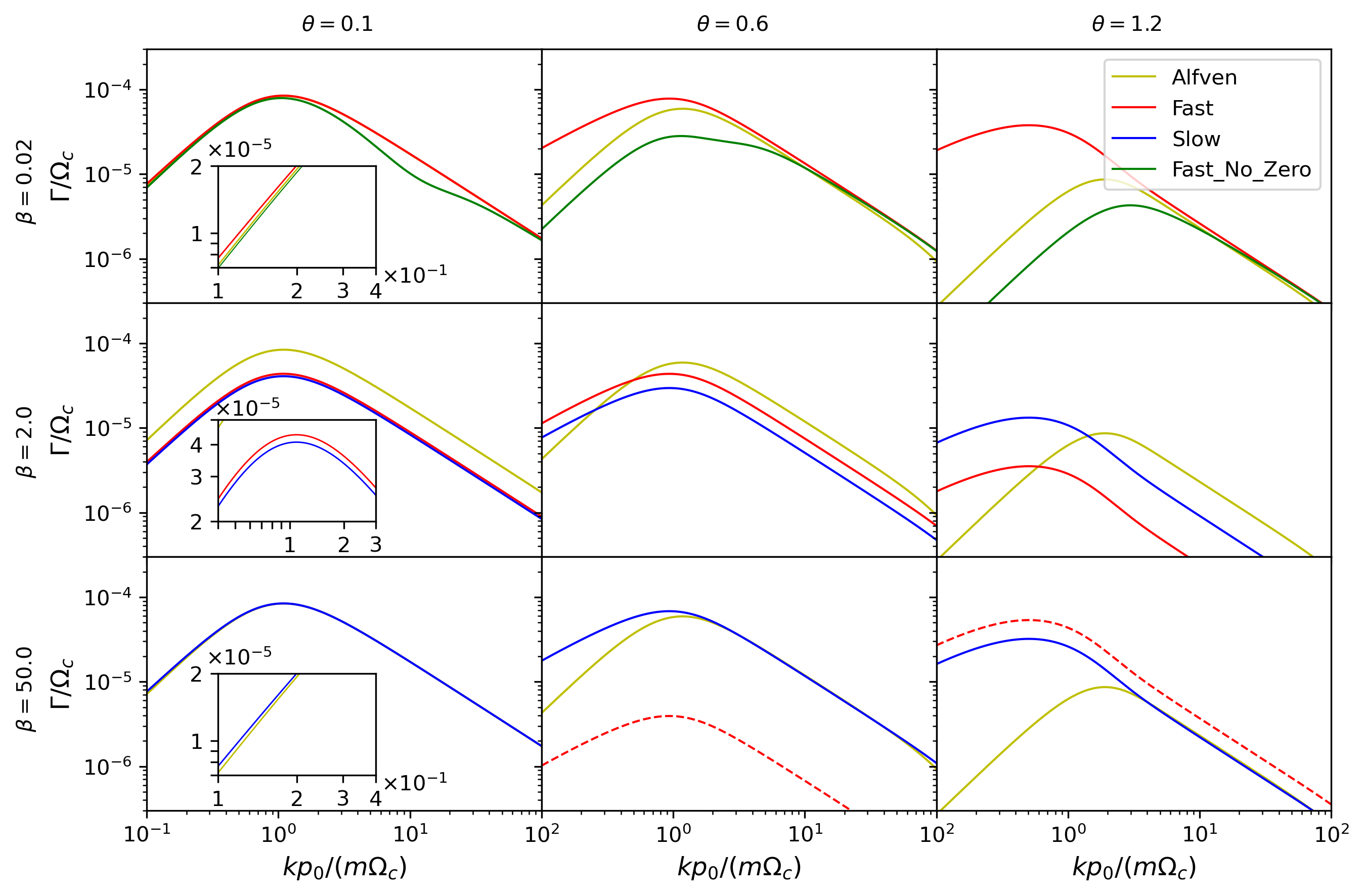}
    \caption{A gallery of the growth or damping rates as a function of wavenumber $k$ with fiducial parameters: $v_d/v_A = 4, \, n_{\mathrm{CR}}/n_i = 10^{-4}$ and $\kappa = 1.25$. The damping rates of the fast waves are shown in red dashed lines. The angle between $\boldsymbol{k}$ and $\boldsymbol{B}_0$, $\theta$, is the same in each column and $\beta$ is the same in each row. We choose $\theta = 0.1, \, 0.6, \, 1.2$ for the first, second and third columns, and $\beta = 0.02, \, 2.0, \, 50$ for each row from top to bottom. In the top panel, we also plot the growth rates of the fast mode excluding the contribution of the zeroth-order resonance in green lines.}
    \label{fig:growth_rate_gallery}
\end{figure*}

In this section, we investigate the general properties of the dispersion relation. For numerical calculations here, we adopt fiducially $v_d/v_A = 4, \, n_{\mathrm{CR}}/n_i = 10^{-4}$. These parameters retains the scale separation consistent with reality, i.e., $n_{\mathrm{CR}}/n_i \ll 1, \, v_A < v_d \ll c$, thus are representative. We choose $\kappa = 1.25$ for the $\kappa$ distribution, such that when $p \gg p_0$, $F(p) \propto p^{-4.5}$. In Figure~\ref{fig:growth_rate_gallery}, we present a gallery of the growth rates of all three modes for the selected parameters. 

Our main findings are summarized as follows:
\begin{itemize}
    \item Plasma $\beta$, which represents the relative importance of thermal energy to magnetic energy, determines the ratio of electric field energy to total wave energy in an MHD wave. The growth rates are proportional to this ratio and thus depend on plasma $\beta$. 
    \item Although the growth rate reaches maximum when $\boldsymbol{k}$ is parallel to $\boldsymbol{B}_0$, we find that it decreases only marginally until the wave vector becomes sufficiently oblique.
    \item When $\theta$ is large, the maximum growth rates of the fast/slow modes can be much higher than that of the Alfven mode. Besides, the most unstable wavenumber roughly follows $1/\cos\theta$ for the Alfven mode, while for the magnetosonic modes it roughly follows $1/\sin\theta$. We attribute these differences to the zeroth-order resonance, the $n=0$ term in the summation in Equation~(\ref{eq:grow rate final}), which is also known as Landau resonance or transit time damping (TTD).
\end{itemize}

\subsection{The Role of Plasma $\beta$}
\label{sec:beta}
The effects of plasma $\beta$ are included in $(k_\parallel v_d/\omega - 1), \, (k_\parallel v_d/\omega - 1)\cos^2\alpha, \, (k_\parallel v_d/\omega - 1)\sin^2\alpha$ for the Alfven, fast and slow modes. Since MHD waves are non-dispersive, the three terms above are all independent of wavelength, meaning that plasma $\beta$ affects the growth rates in the same way for all wavelengths. Therefore, the shape of the growth rate curves in each column of Figure~\ref{fig:growth_rate_gallery} remains unchanged with varying $\beta$. A direct inference is that the most unstable wavenumber, $k_m$, is independent of $\beta$. In addition, one can see that the magnetosonic modes share the same $k_m$ since they have the same integrand in Equation~(\ref{eq:final}). For the Alfven mode, we have $(k_\parallel v_d/\omega - 1) = (v_d/v_A - 1)$ being independent of $\beta$, meaning that this mode is not affected by plasma $\beta$. This is reasonable since the Alfven mode is incompressible and does not "feel" the thermal pressure. Therefore, we focus on the magnetosonic modes in the following.

Equation~(\ref{eq:cos_sin_alpha}) shows that plasma $\beta$ determines the ratio of electric field energy to total wave energy in an MHD wave, to which the growth rates are proportional. With increasing $\beta$, fast waves become increasingly acoustic and slow waves become increasingly electromagnetic. Since sound waves do not involve electromagnetic field fluctuations, they do not exchange energy with cosmic rays. Therefore, with increasing $\beta$, the growth rate of fast waves decreases while the growth rate of slow waves increases, which is shown in each column of Figure~\ref{fig:growth_rate_gallery}. When $\beta$ exceeds $\beta_c \equiv 2(v_d/v_A)^2$, the sound speed surpasses the drift speed, thus fast waves are damped regardless of the propagation direction.

\subsection{Propagation Direction Dependence}
\label{sec:angle dependence}
\begin{figure}
    \includegraphics[width=\columnwidth]{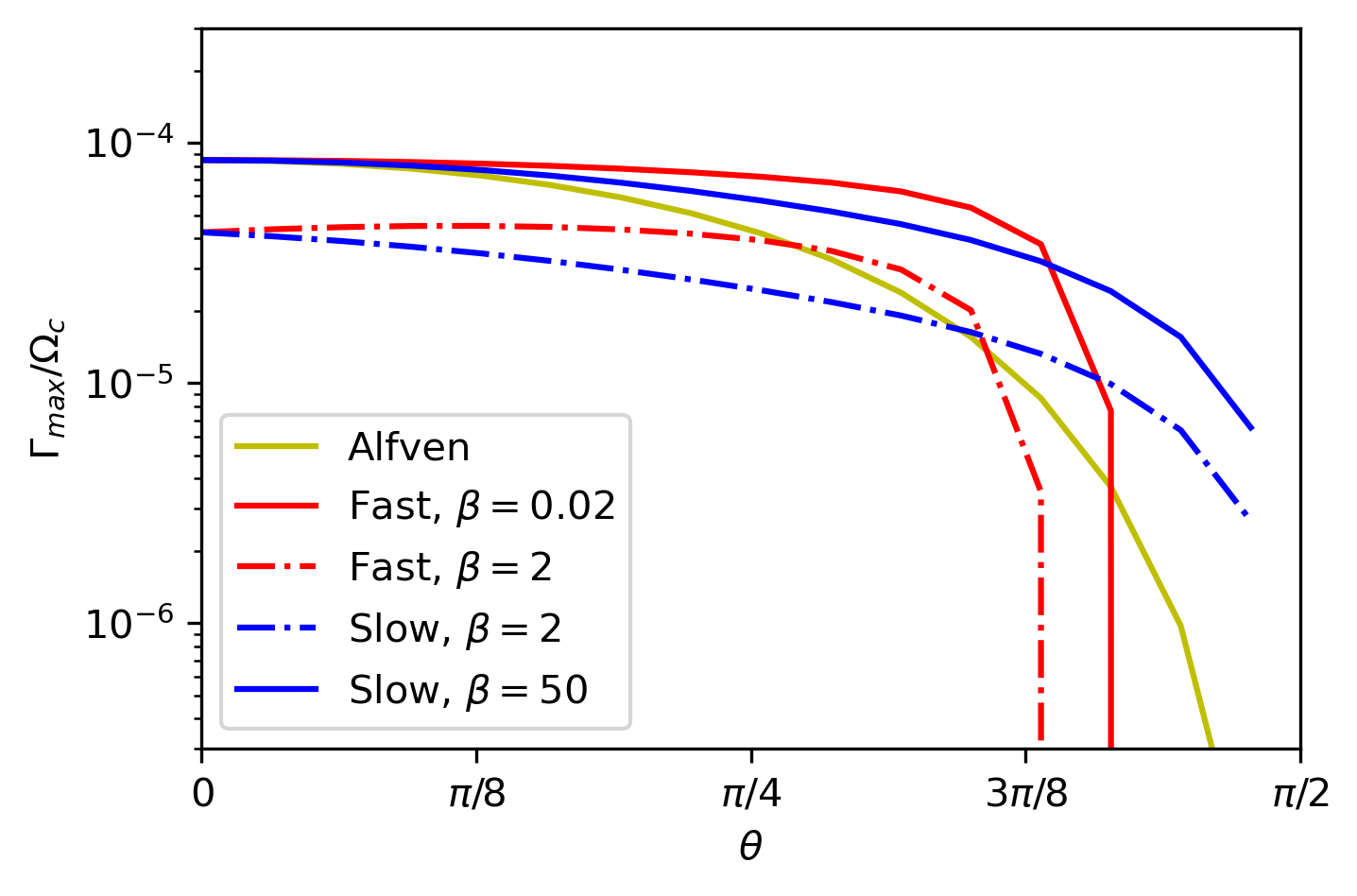}
    \caption{Maximum growth rates of each mode as a function of $\theta$ at different $\beta$s. The growth rates of Alfven waves do not depend on $\beta$. When $\beta=50$, fast waves are damped, thus their (negative) growth rates are not shown in the figure. When $\beta=0.02$, the growth rates of slow waves are too small to appear in the figure.}
    \label{fig:max_growth_rate}
\end{figure}
\begin{figure}
    \includegraphics[width=\columnwidth]{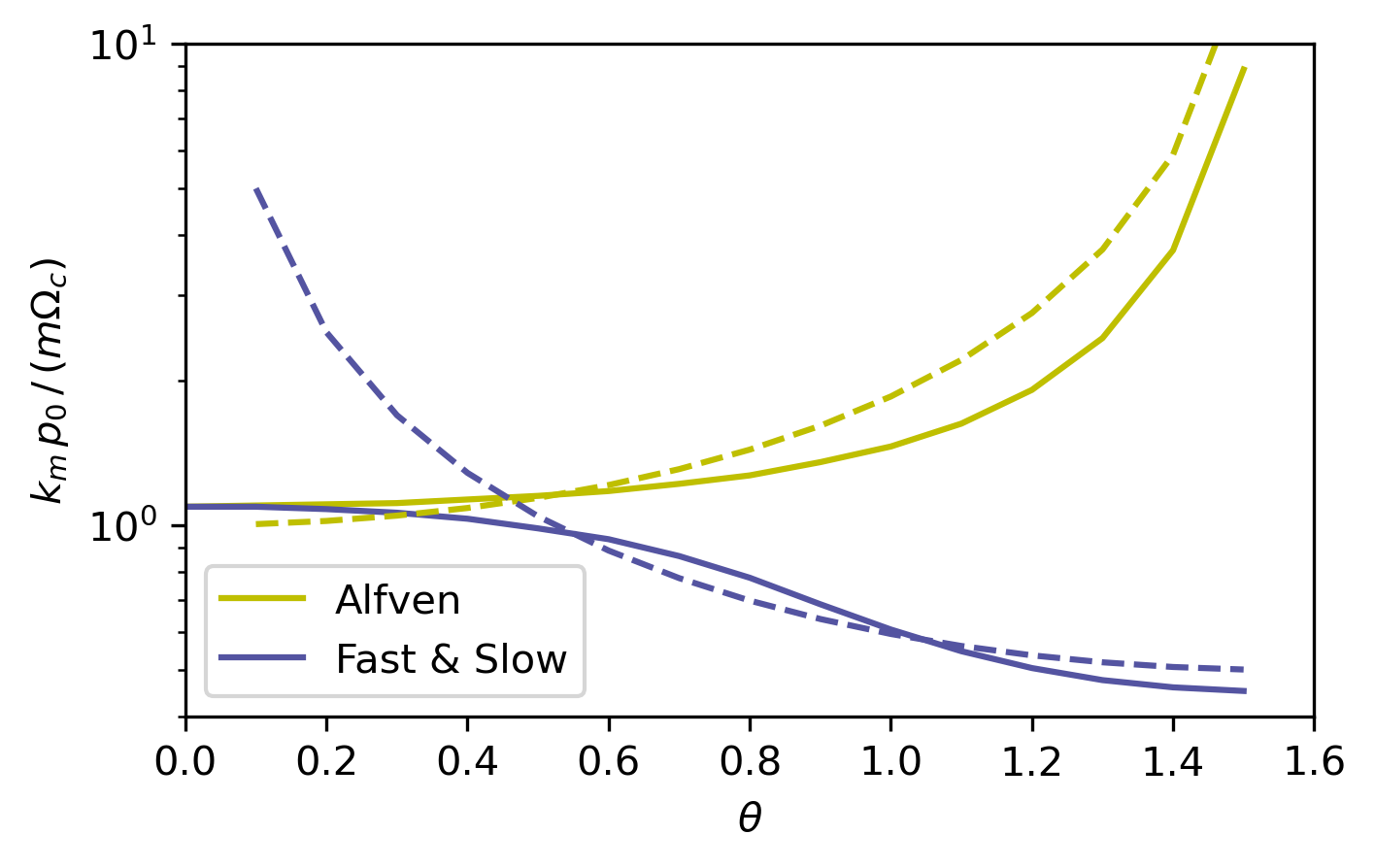}
    \caption{The most unstable wavenumbers $k_m$ as a function of $\theta$ for different wave modes (note that $k_m$ is independent of $\beta$ as discussed in Section~\ref{sec:beta}). The dashed lines in yellow and purple mark the wavenumbers with constant $k_\parallel$ and $k_\perp$ respectively (the yellow line shows $k_m p_0/(m\Omega_c) =  1/\cos \theta$ and the purple line shows $k_m p_0/(m\Omega_c) = 1/(2\sin \theta)$), which reasonably well reproduce the trends for $\theta > 0.6$.}
    \label{fig:most_unstable_k}
\end{figure}
We note that the growth rates are similar across a wide range of $\theta$ values. The first and second columns in Figure~\ref{fig:growth_rate_gallery} show that the growth rates of the dominant branch are comparable across the entire spectrum when comparing $\theta = 0.1$ and $\theta = 0.6$ cases. Low-$k$ magnetosonic waves even grow faster when $\theta = 0.6$. In Figure~\ref{fig:max_growth_rate}, we plot the maximum growth rates, $\Gamma_{\mathrm{max}}$, of each mode as a function of $\theta$ at different $\beta$s. Although, in general, the maximum growth rates decrease with increasing $\theta$ (except for the slowest-growing one among the three, whose growth rate is negligible), this decrease is modest until $\theta$ becomes large, $\theta > \pi/4$ \citep{Tademaru_1969}. These analyses indicate that the growth of oblique waves and their interaction with CRs are non-negligible. In Figure~\ref{fig:max_growth_rate}, we can see the existence of a critical angle, $\theta_c$, beyond which the fast waves are damped. The explicit form of $\theta_c$ can be derived from $(k_\parallel v_d/\omega - 1)=0$:
\begin{equation}
    \cos^2 \theta_c = (\frac{v_A}{v_d})^2 \, \bigg[ 1 + \frac{\beta}{2}\big(1 - (\frac{v_A}{v_d})^2\big)\bigg].
    \label{eq:theta_c}
\end{equation}
where we have used the dispersion relation for the fast waves (Equation~\ref{eq:MHD wave}).

In Figure~\ref{fig:most_unstable_k}, we plot the most unstable wavenumbers $k_m$ as a function of $\theta$ for different wave modes. For Alfven and Slow modes, these wavenumbers correspond to fastest growth, while for the fast mode, they correspond to either fastest growth or fastest damping. When $\theta>0.6$, the most unstable wavenumber of the Alfven waves roughly follows $k_m \propto 1/\cos\theta$, while that of the magnetosonic waves roughly follows $k_m \propto 1/\sin\theta$. We will discuss the origin of this difference in the next section.

\subsection{Zeroth-order Resonance}
\label{sec: zeroth-order}
In Figure~\ref{fig:growth_rate_gallery}, we see that when $\theta = 1.2$, growth rates of the magnetosonic modes can be much higher than those of the Alfven mode in the low-$k$ region. Furthermore, the behavior of the most unstable wavenumbers of the Alfven mode differs significantly from that of the magnetosonic modes when $\theta$ is large, as shown in Figure~\ref{fig:most_unstable_k}. We find that these differences are caused by the zeroth-order resonance. In Equation~(\ref{eq:grow rate final}), it is straightforward to see that the $n=0$ term is nonzero for magnotosonic modes but zero for the Alfven mode. 

In the top panel of Figure~\ref{fig:growth_rate_gallery}, the green lines show the growth rates of the fast mode excluding the contribution of the zeroth-order resonance. By comparing the red and green lines, one can see that the zeroth-order resonance has little contribution when $\theta = 0.1$, but dominates the growth of the fast wave when $\theta = 1.2$. The same analysis also applies to the slow mode. While the zeroth-order resonance of magnetosonic waves has been well studied in the context of cosmic-ray transport with energy gain \citep{Xu&Lazarian_2018, Schlickeiser_1998}, here we study its role in detail in the excitation of the CRSI (and hence CR energy loss).

The zeroth-order resonance can be best understood in the wave frame, which moves with a velocity $u^w = \omega/k_\parallel$ along the background field $\boldsymbol{B_0}$. Since the three wave modes have different phase speeds, there are, in principle, three wave frames. However, in our system, the dominant magnetosonic modes are near-parallel Alfven-like. Thus, we take $u^w = v_A$ hereafter, which is exact for the Alfven mode and serves as a good approximation for the dominant magnetosonic modes. In this frame, the electric field vanishes, and the perturbed magnetic field, $\delta \boldsymbol{B}^w$, reads,
\begin{equation}
    \delta \boldsymbol{B}^w = \big[\delta B_a \Hat{y} + (\delta B_f + \delta B_s) (\cos\theta \Hat{x} - \sin\theta \Hat{z})\big] \exp{\big[\mathrm{i} (\boldsymbol{k} \cdot \boldsymbol{r^w})\big]},
\end{equation}
where the superscript $w$ refers to the wave frame and we have neglected the difference between $\delta B_i^w$, $\boldsymbol{k}^w$ and $\delta B_i$, $\boldsymbol{k}$, since $u^w \ll c$. To the first-order of $\delta B/B_0$, the energy density of magnetic field is
\begin{equation}
    \mathscr{E}_b = \frac{B_0^2}{8\pi} \bigg[1 - 2 \sin\theta \, \frac{(\delta B_f + \delta B_s)}{B_0} \cos(\boldsymbol{k} \cdot \boldsymbol{r^w}) \bigg].
    \label{eq:magnetic energy in wave frame}
\end{equation}
The fluctuation part comes only from obliquely propagating magnetosonic waves since they can create longitudinal magnetic field perturbation, whereas the Alfven wave can not. These strong perturbations in magnetic field energy density can reflect particles in a way similar to magnetic mirrors, thus we refer to them as mirror-like structures in this paper. The typical scale of the perturbed magnetic field is $\lambda_\parallel = 2\pi / k_\parallel$ in the parallel direction and $\lambda_\perp = 2\pi / k_\perp$ in the perpendicular direction. For particles with $v_\parallel^w \sim c$, they can pass through multiple $\lambda_\parallel$ within one gyro-period, so on average the mirror-like structures have no net effect on them. For particles with $v_\parallel^w \sim 0$, the distance they travel in multiple gyration periods is less than $\lambda_\parallel$, thus their motion is strongly influenced by the mirror-like structures. This analysis is consistent with the condition for the zeroth-order resonance, $v_\parallel^w = 0$. In the adiabatic approximation, the longitudinal motion is determined by
\begin{equation}
    \frac{\mathrm{d} v_\parallel}{\mathrm{d}t} = - 2\pi \frac{v_\perp^2}{B_0^2} \frac{\mathrm{d} \mathscr{E}_b}{\mathrm{d} z} = - \frac{v_\perp^2}{4} k \sin 2\theta \frac{(\delta B_f + \delta B_s)}{B_0} \sin(\boldsymbol{k} \cdot \boldsymbol{r^w}).
    \label{eq:mirror_force}
\end{equation}
By applying a longitudinal force, the mirror-like structures can reflect particles with pitch angles sufficiently close to 90-degree. 

The adiabatic approximation requires that the magnetic field experienced by particles does not change significantly within one gyration motion. This assumption holds only when $k_\perp r_L \lesssim 1$, where $r_L$ is the gyroradius of the particle. Within this range, Equation~(\ref{eq:mirror_force}) shows that the strength of the parallel force increases as $k$ increases. When $k_\perp r_L \gg 1$, particles cross multiple $\lambda_\perp$ within one gyration period, so there is no net effect of the mirror-like structures. This is consistent with the fact that the red and green lines overlap in the high-$k$ region in Figure~\ref{fig:growth_rate_gallery}, which means the zeroth-order resonance has little effect on short-wavelength waves. Combining these two conditions, it can be concluded that for the zeroth-order resonance, the most unstable wavenumber is given by $(k_m)_\perp r_L \sim 1$. For the $\kappa$ distribution, its derivative reaches the maximum at $p \sim p_0$. These particles make the most significant contribution to wave growth. Therefore, the typical $r_L$ can be estimated as $r_L \sim p_0/(m\Omega_c)$. Then, for the case dominated by the zeroth-order resonance, the most unstable wavenumber roughly follows $(k_m)_\perp p_0 / (m\Omega_c) \sim 1$. This explains the behavior of $k_m \propto (\sin\theta)^{-1}$ for magnetosonic waves when $\theta$ is large, as shown in Figure~\ref{fig:most_unstable_k}. For comparison, the growth of Alfven waves is solely supported by cyclotron resonances ($n \neq 0$), so the most unstable wavenumber roughly follows $(k_m)_\parallel p_0 / (m\Omega_c) \sim 1$, meaning that $k_m \propto (\cos\theta)^{-1}$.

\section{Simulation Setup and Choice of Parameters}
\label{sec:simulation setup}
\begin{table*}
	\centering
	\caption{List of Main Simulation Runs}
	\label{tab:main_runs}
	\begin{tabular}{c@{\hskip 10pt}c@{\hskip 10pt}c@{\hskip 10pt}c@{\hskip 10pt}c@{\hskip 10pt}c@{\hskip 10pt}c@{\hskip 10pt}c@{\hskip 10pt}c@{\hskip 10pt}c@{\hskip 10pt}c@{\hskip 10pt}c@{\hskip 10pt}c}
		\hline
		 Run  & dimension &$\theta$ & $\beta$        & $v_d/v_A$ & $n_{\mathrm{CR}}/n_i$ & $L_x(d_i)$ & $L_y(d_i)$ & $\Delta x(d_i)$ & $\Delta y(d_i)$ & $A_0$ & $N_p$ & t ($\Omega_c^{-1}$) \\
		\hline
          1D-s  & 1D & 0.1/0.6 & 0.02/2/50 & 4 & $1\times 10^{-4}$ & 4,800,000 & \text{--} & 10 & \text{--} & $1.0\times10^{-5}$ & 256 & $1 \times 10^4$  \\
          1D-l  & 1D & 1.2 & 0.02/2/50 & 4 & $1\times 10^{-4}$ & 9,600,000 & \text{--} & 10 & \text{--} & $1.0\times10^{-5}$ & 256 & $1 \times 10^4$  \\
          2D  & 2D & \text{--} & 0.02/2/50 & 4 & $3 \times 10^{-4}$ & 36,000 & 18,000 & 15 & 15 & $1.0\times10^{-5}$ &16 & $1 \times 10^5$  \\
          1D-c  & 1D & 0 & 0.02 & 4 & $3 \times 10^{-4}$ & 36,000 & \text{--} & 15 & \text{--} & $3.7\times10^{-5}$ & 256 & $1 \times 10^5$  \\
		\hline
	\end{tabular}
\end{table*}

We study the linear growth of the CRSI and later quasi-linear evolution through 1D and 2D MHD-PIC simulations. We use the Athena++ MHD code \citep{Stone_2020} with the MHD-PIC module implemented by \cite{Sun_2023}. Our simulation setup mostly follows \cite{Bai2019} with only minor changes. In Table~\ref{tab:main_runs}, we summarize all simulation runs presented in this paper with the chosen parameters. 

We set up the simulation in the drift frame, where the CR momentum distribution is isotropic, $F(\boldsymbol{p}) = F(p)$. The distribution function is chosen to be a $\kappa$ distribution with $\kappa = 1.25$, $p_0 = 300m v_A$. To ensure that the entire distribution function is well resolved, we divide the momentum space into eight bins ranging from $p_0/500$ to $500p_0$ and inject an equal number ($N_p$) of CR particles per cell per bin. In this frame, the gas has a bulk motion with velocity $-\boldsymbol{v_d}$. For simplicity, we adopt the isothermal equation of state with the background density being $\rho_0 = 1$. The plasma $\beta$ is chosen to be $0.02, 2, 50$, same as in previous analytical calculation. The magnitude of the background magnetic field is set to $B_0 = 1$, and the Alfven speed is then $v_A = B_0/\sqrt{\rho_0} = 1$. The code units are chosen such that the cyclotron frequency $\Omega_c = 1$ for CRs in the background field $\boldsymbol{B_0}$, and a natural unit of time is $\Omega_c^{-1}$. The length unit in the simulation is then $d_i = v_A/\Omega_c = 1$. The artificial speed of light is chosen to be $\mathbb{C} = 300v_A$.

In our 1D simulations, we align the simulation box with the wave vector, and there is an angle $\theta$ between the background magnetic field and the simulation box. On top of the bulk motion of gas, we initialize the system with a spectrum of waves. For each wavenumber $|k|$, there are six wave modes: forward- and backward-propagating Alfven waves, fast waves, and slow waves. The wavenumbers cover the range from $|k| = 2\pi/L$ to $|k| = 2\pi/(2\Delta x)$ (except that the shortest-wavelength wave is initialized with zero amplitude), where $L = N_x \Delta x$ is the simulation domain size, $N_x$ and $\Delta x$ are the number of grid cells and cell size. Correspondingly, we set up $6(N_x/2-1)$ modes in total. The eigenvector of each mode, $\boldsymbol{R}_i = (\rho_1, \boldsymbol{u}_1, \boldsymbol{B}_1)$ can be obtained by solving the linearized MHD equations, and its explicit expression is provided in Appendix \ref{appen C}. We normalize $\boldsymbol{R}_i$ such that $|\boldsymbol{u}_1| = v_A$. The perturbed quantities in the simulations are set by $A(k) \boldsymbol{R}_i \sin(kx+\psi_i(k))$, where $A(k)$ is the wave amplitude, $x$ is the coordinate along the simulation box, and $\psi$ is a random phase. The form of $A(k)$ is chosen to be $A(k) = A_0/\sqrt{k}$ with $A_0$ being a constant, and energy density of a single wave mode is $I(k) = (A_0^2/k)(B_0^2 / 2)$, which is equally distributed in logarithmic $k$-space. The total wave energy density is 
\begin{equation}
    \frac{I_{\mathrm{tot}}}{B_0^2 / 2} = 6 A_0^2 \ln (\frac{N_x}{2} - 1).
\end{equation}

We perform simulations for $\theta = 0.1, 0.6, 1.2$, same as in previous analytical discussion. When $\theta = 0.1, 0.6$, the most unstable wavelength is near $\lambda_0 = 2\pi p_0/(m\Omega_c) = 1885 d_i$ for three modes, as shown in Figure~\ref{fig:most_unstable_k}. Thus, in these two cases, we choose the simulation box size $L$ to be $4800000 d_i$, $\sim 2500$ times the most unstable wavelength. When $\theta = 1.2$, the most unstable wavelength of the Alfven mode is about $\lambda_0/2$, while for the magnetosonic modes it is about $2\lambda_0$. Therefore, we set $L = 9600000 d_i$ to maintain it approximately $2500$ times the longer most unstable wavelength. The resolution ($d_i$ per cell) is set to $\Delta x = 10 d_i$.

In our 2D simulations, the background field $\boldsymbol{B}_0$ is set along $\hat{x}$, and the simulation is performed in the $x\text{--} y$ plane. Similar to the 1D simulations, we initialize a spectrum of waves, with parallel wavenumbers $|k_x|$ ranging from $2\pi/L_x$ to $2\pi / (2\Delta x)$ and perpendicular wavenumbers $|k_y|$ ranging from $2\pi/L_y$ to $2\pi / (2\Delta y)$ (waves with the maximum wavenumbers have their amplitudes set to zero). For a given $(|k_x|, |k_y|)$, a total of 12 wave modes are initialized: Alfven waves, fast waves, and slow waves, each propagating along four directions corresponding to the four quadrants of $\boldsymbol{k}$-space. Additionally, a series of waves propagating along the background magnetic field ($k_y = 0$) are added in the simulation, and their setup is identical to that in the 1D simulations with $\theta = 0$. In this setup, a total of $6(N_x/2-1)[(2(N_y/2-1) + 1]$ waves are added in the simulation. For waves with $k_y \neq 0$, their amplitudes are determined by $A(k_x, k_y) = A_0/\sqrt{k_x k_y}$.
The total wave energy density is 
\begin{equation}
    \frac{I_{\mathrm{tot}}}{B_0^2 / 2} = 6 A_0^2 \ln (\frac{N_x}{2} - 1) \bigg[2\ln (\frac{N_y}{2} - 1) + 1\bigg].
\end{equation}

The size of the simulation box is $L_x = 36000 d_i, L_y = 18000 d_i$, with a resolution of $\Delta x = \Delta y = 15 d_i$. This relatively small simulation box is accompanied with the phase randomization procedure described in \cite{Bai2019}. The basic idea is to rotate the momentum of each particle around the background field by a random angle every $L_x/\mathbb{C}$. This allows particles to effectively see different wave packets throughout the simulation. In 1D-s and 1D-l runs, since the computational cost is manageable, we choose to use a longer simulation box instead of employing phase randomization.

In both 1D and 2D simulations, the initial amplitude is set to $A_0 = 10^{-5}$ to keep the system in the linear stage for a relatively long time, allowing for more accurate measurement of growth rates. The drift speed is chosen as $v_d=4v_A$. For 1D runs, we use $n_{\rm CR}/n_i = 1 \times 10^{-4}$, while in 2D we increase it to $3 \times 10^{-4}$, which leads to faster growth and earlier saturation, thus reducing computational cost. The number of particles per cell per bin is $N_p=256$ in 1D runs and $N_p=16$ in 2D runs.

To better investigate the effects of oblique waves, we conduct a 1D simulation (run 1D-c) containing only parallel waves. Its setup is identical to that in the 1D simulations with $\theta = 0$, except that the phase randomization procedure is employed here to ensure consistency with the 2D simulations. The parameters are the same as those in 2D runs, except for the initial amplitude and number of particles per cell. We set $A_0 = 3.7 \times 10^{-5}$ for run 1D-c such that the initial energy density of waves is identical in run 1D-c and 2D runs. The number of particles per cell per bin is chosen as $N_p = 256$ to ensure sufficient resolution in momentum space of the CRs.

\begin{figure*}
    \includegraphics[width=17cm]{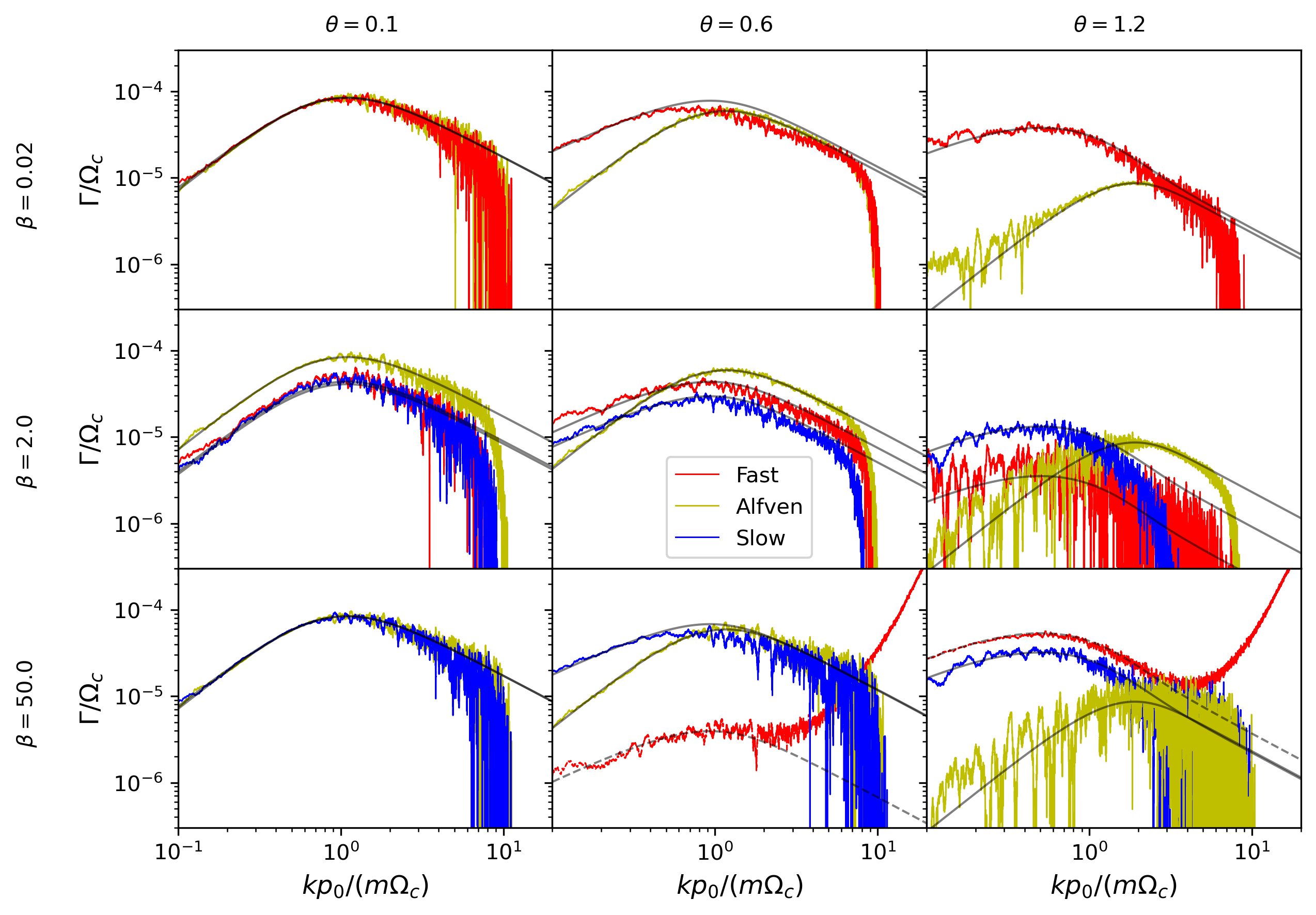}
    \caption{Growth and damping rates measured from 1D simulations with theoretical comparison. Solid and dashed lines represent growth and damping rates, respectively. Yellow, red and blue lines correspond to the measured growth/damping rates of the Alfven, fast, and slow modes in the simulations.  Black lines are analytical values. The raised ends of the red dashed lines in the last two panels are due to the numerical dissipation.}
    \label{fig:1d_sim}
\end{figure*}

\section{Simulation Results}
\label{sec:sim results}
\subsection{Growth Rate in 1D Simulations}
\label{sec:1d sim results}
We measure the growth and damping rates of three forward-propagating wave modes from simulations using the method described below. At a sequence of evenly spaced time steps, we decompose the perturbed fields into six modes and obtain $I(k, t)$ for each mode (see Appendix~\ref{appen C}). To reduce noise, for each time step, we apply a moving average filter with a window width of $\Delta k = 100 \times (2\pi/L)$ (that is, 100 data points) to the spectra $I(k, t)$. A linear fit is then performed on $\ln[I(k, t)]$, with the slope being $2\Gamma(k)$. The entire time interval for measuring growth rates should not be too short, otherwise even the fastest-growing mode can not grow much, making it difficult to measure growth rates accurately. The interval should also not be too long, otherwise the quasi-linear diffusion (QLD) would start to modify the CR distribution function. We adopt the time interval being $t \leq 5\times 10^3 \Omega_c^{-1}$ except for $\beta = 2, \theta=1.2$ case, in which we adopt $t \leq 1\times 10^4 \Omega_c^{-1}$ since the maximum growth rates are relatively small. 

The measured growth and damping rates are shown in Figure~\ref{fig:1d_sim}. Under all parameter conditions, the growth and damping rates measured in the simulations well match the theoretical predictions across a wide range of wavenumbers. The growth and damping of the CRSI are better captured in the low-$k$ regime. In the high-$k$ regime, the CRSI growth competes with numerical dissipation, leading to a cutoff in the measured growth rates at $k_{\rm cut}\sim 10 m\Omega_c / p_0$. In the panel for $\beta = 50$ and $\theta = 0.6$, the red line matches the black dashed line at small $k$, indicating that the damping of the fast wave is dominated by the CRSI. However, as $k$ approaches $k_{\rm cut}$, numerical dissipation becomes dominant, eventually surpassing the CRSI growth at $k \gtrsim k_{\rm cut}$, resulting in the cutoff. 

\subsection{2D Simulation}
\label{sec:2d sim}
\begin{figure}
    \includegraphics[width=\columnwidth]{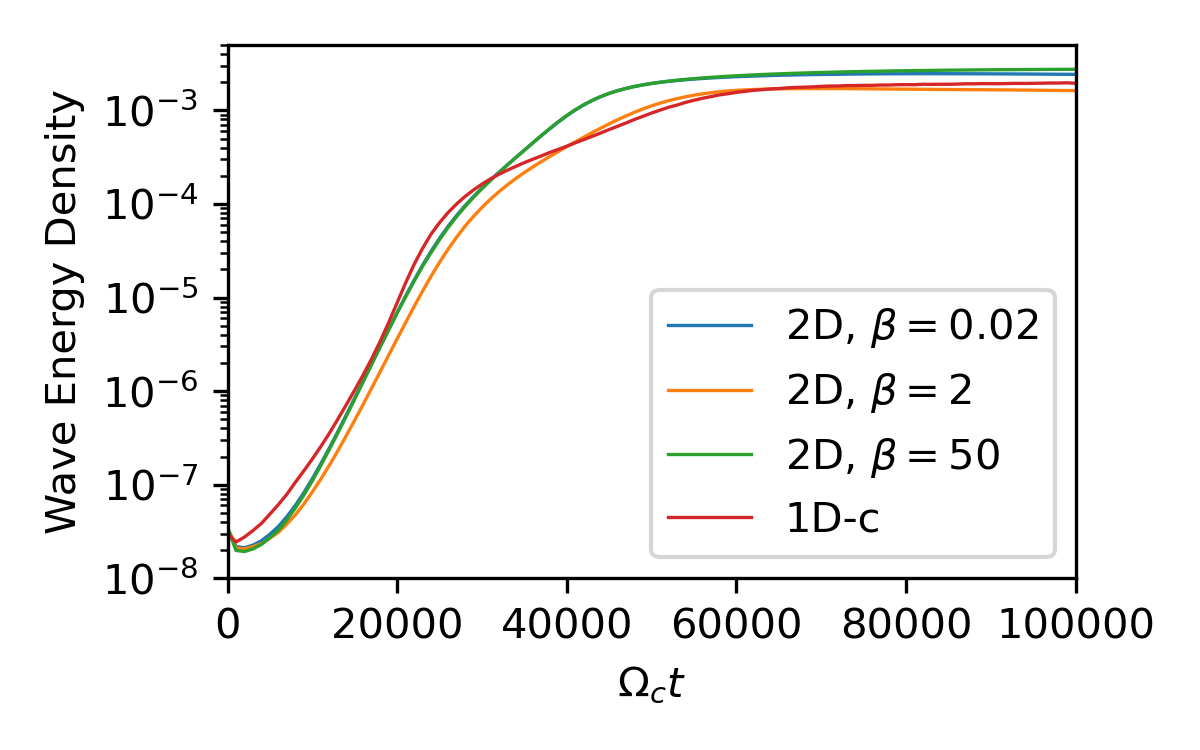}
    \caption{Time evolution of total wave energy density in 2D and 1D-c runs. All numbers are in code units.}
    \label{fig:evo_e}
\end{figure}
\begin{figure*}
    \includegraphics[width=17cm]{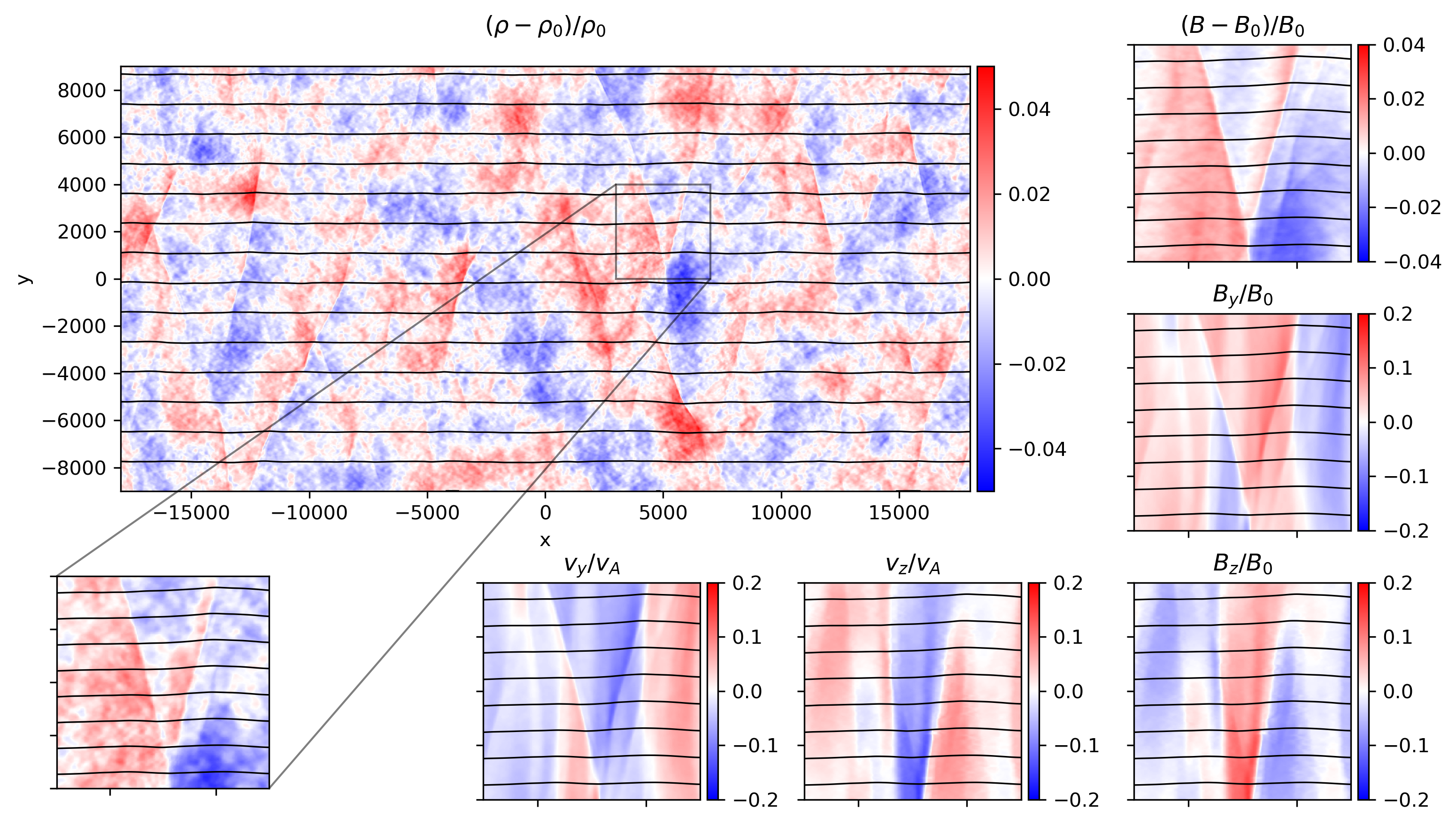}
    \caption{Perturbed density at $t = 1 \times 10^5 \Omega_c^{-1}$ in the 2D run with $\beta = 0.02$, overplotted with the field lines. In the zoom-in panels, we show the perturbed density, perturbed magnetic field strength, and the $y, z$ components of fluid velocity and magnetic field.}
    \label{fig:field_structure}
\end{figure*}
Having successfully reproduced the analytical growth rates of oblique CRSI modes in 1D, we now proceed to 2D simulations. This will allow us to simultaneously measure the growth rates of waves propagating in different directions and study the process of their collective scattering of CRs. Besides, the multi-dimensional nature of these simulations enables us to discuss the role of plasma $\beta$.

In Figure~\ref{fig:evo_e}, we show the growth history of the energy density of waves in all 2D runs and run 1D-c. The energy density of waves in all simulations exhibit similar growth behaviors: they grow exponentially until $t \sim 3 \times 10^4 \Omega_c^{-1}$, then transition into a quasi-linear evolution stage, and finally reach comparable saturated values after $t \sim 6 \times 10^4 \Omega_c^{-1}$. Despite the similarity, there are slight differences: waves in run 1D-c grow faster at first since parallel modes on average have larger linear growth rates, but saturate at a slightly lower level compared with 2D runs with $\beta=0.02, 50$. Besides, in run 1D-c, there is a knee-like feature at $t \sim 2.5 \times 10^4 \Omega_c^{-1}$, where a fast growth phase transitions to a slow growth phase. In contrast, the transition from fast growth to saturation in the 2D simulations is smoother. 

In Figure~\ref{fig:field_structure}, we show the perturbed density of the entire simulation domain in saturated state ($t = 1 \times 10^5\Omega_c^{-1}$) together with the field lines. The perturbed magnetic field strength, and the $y, z$ components of fluid velocity and magnetic field are shown in the zoom-in panels. Here, we choose $\beta = 0.02$ as a representative, but the following features are also found in the other two 2D runs:
\begin{itemize}
    \item The field lines wiggle thanks to the presence of waves, while neighboring field lines are largely parallel to each other, suggesting that the waves remain quasi-parallel. No clear magnetic mirror/bottle configuration is observed.
    \item There are oblique Alfven and magnetosonic waves (in this case, fast waves) sweeping the entire domain. These waves can be identified through the correlations between perturbed fields: $v_z/v_A \approx -B_z/B_0$ matches the eigenvector of the Alfven mode, $v_y/v_A \approx -B_y/B_0$ matches the eigenvector of the fast mode in the low-$\beta$ limit. Furthermore, the correlation between $(\rho - \rho_0)/\rho_0$ and $(B-B_0)/B_0$ suggests that the field strength perturbation is primarily driven by compressible oblique magnetosonic waves, consistent with the analysis in Section~\ref{sec: zeroth-order}.
    \item These waves interfere with each other and form weak shocks, i.e., abrupt change of sign of physical quantities in a short spatial scale. We note that there are regions where $B_y$ and $B_z$ change sign, whereas $B$ remains approximately constant. They may be the counterparts of the rotational discontinuities found in 1D simulations \citep{Plotnikov_2021}.
\end{itemize}

In the following subsections, we first present the measured growth rates during the linear phase. Next, we discuss the quasi-linear evolution and the saturated state. We end this section with an investigation of how particles overcome the well-known 90-degree barrier.

\subsubsection{Linear Wave Growth}
\begin{figure}
    \includegraphics[width=\columnwidth]{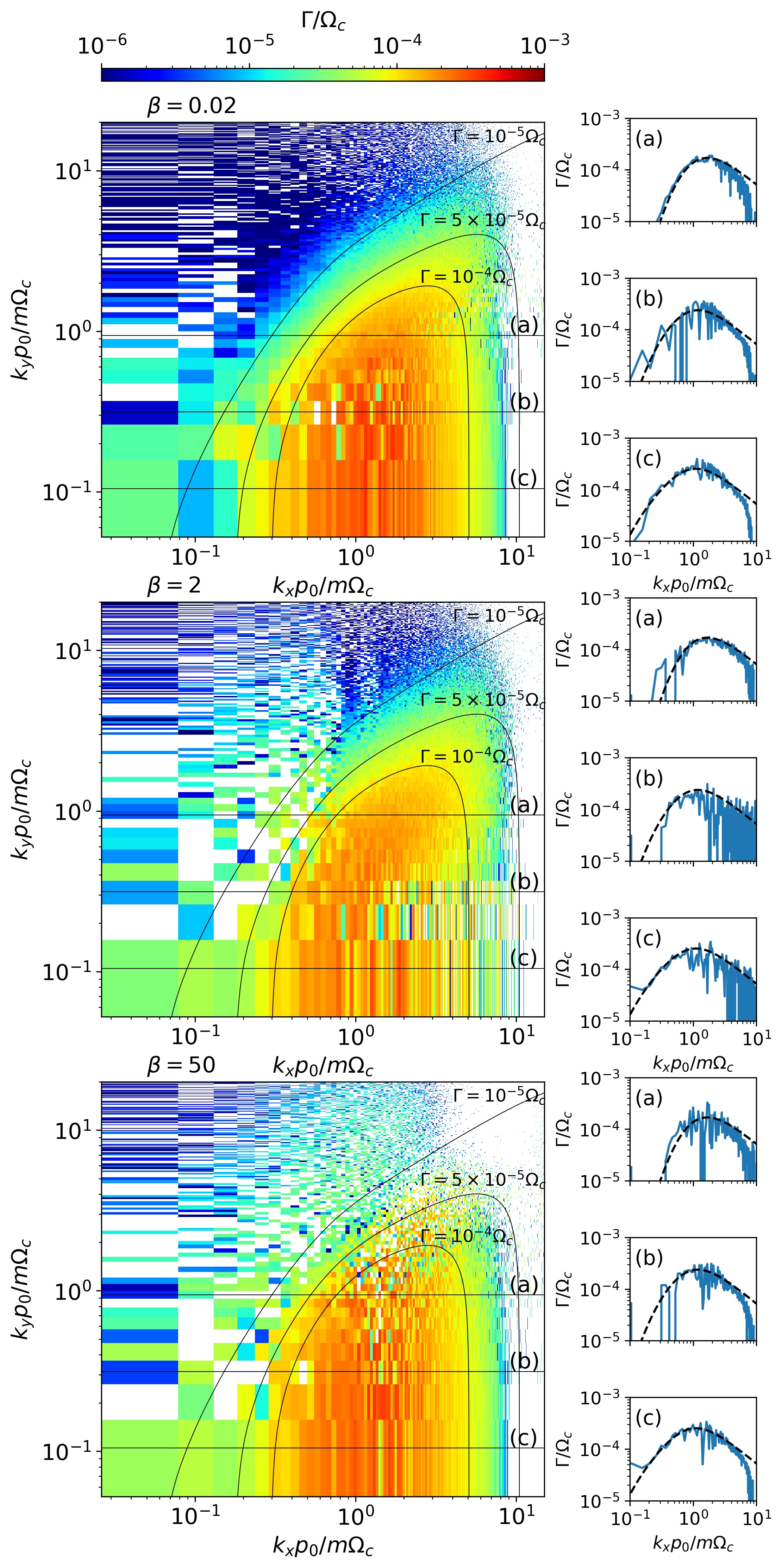}
    \caption{Growth rates of Alfven waves measured from 2D simulations with $\beta = 0.02, 2, 50$. Since the growth rates is symmetric about $k_x$, we only show the growth rates of waves in the first quadrant of the $\boldsymbol{k}$-space. In the left panels, the black contours are analytical predictions. To better compare the simulation and theory, we select three slices at $k_y = 2\pi/L_y, 6\pi/L_y$ and $18\pi/L_y$. They are represented by the straight lines in each of the three left panels, and we plot the analytical (black dashed) and simulation (blue solid) results at these slices in the right panels.}
    \label{fig:2d_alfven}
\end{figure}
\begin{figure}
    \includegraphics[width=\columnwidth]{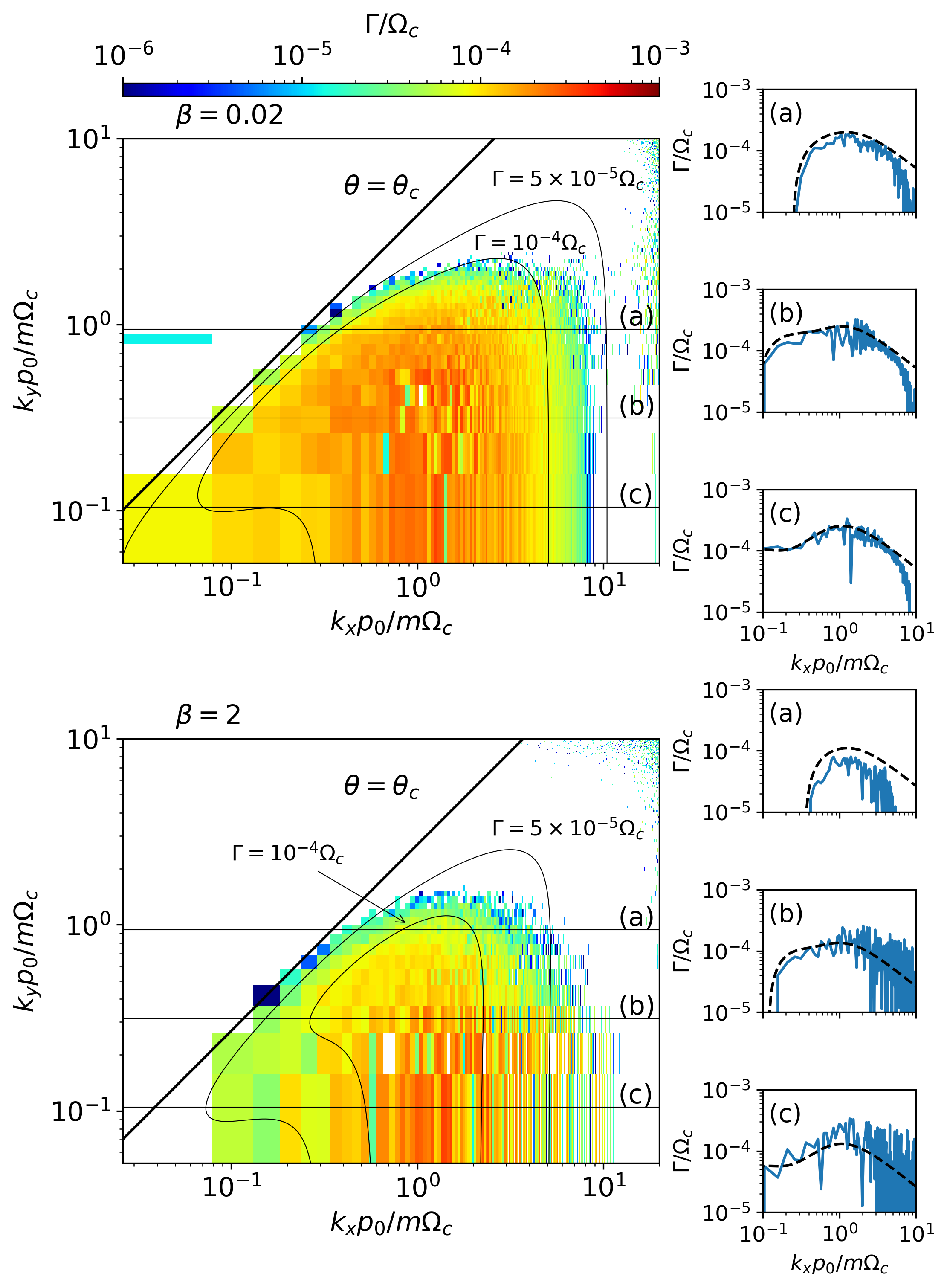}
    \caption{Same as Figure~\ref{fig:2d_alfven} but for fast waves with $\beta = 0.02, 2$. When $\beta = 50$, fast waves decay rather than grows, thus this case is not shown here. We also plot the $\theta_c$, beyond which fast waves decays.}
    \label{fig:2d_fast}
\end{figure}
\begin{figure}
    \includegraphics[width=\columnwidth]{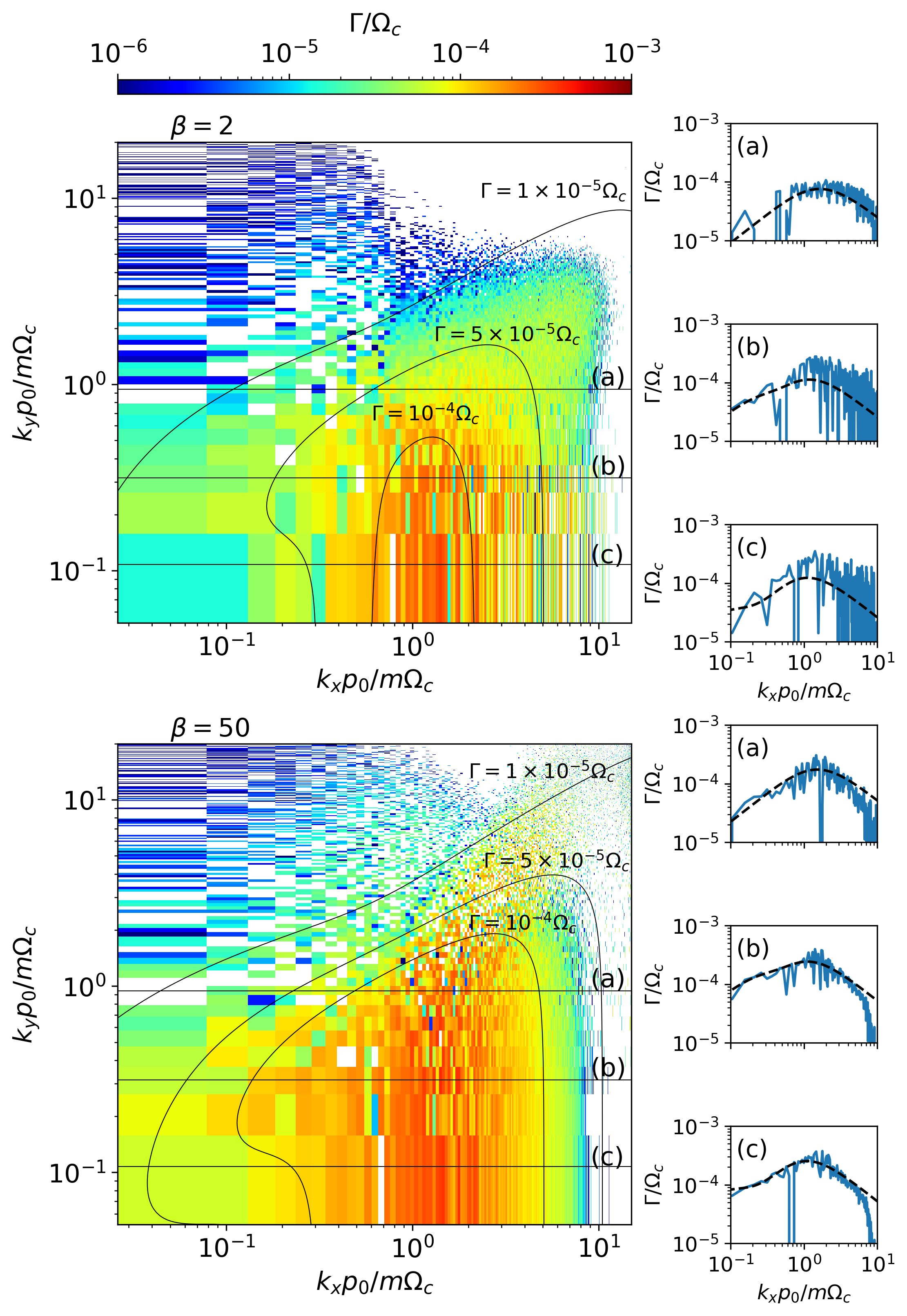}
    \caption{Same as Figure~\ref{fig:2d_alfven} but for slow waves with $\beta = 2, 50$. When $\beta = 0.02$, growth rates of slow waves is negligible, thus this case is not shown here.}
    \label{fig:2d_slow}
\end{figure}
We measure the linear growth rates of forward-propagating waves from 2D simulations based on a similar procedure as described in Section~\ref{sec:1d sim results}. The 2D box is approximately 100 times shorter than the 1D box, and the resolution in $k$-space is comparable to the window width of the filter adopted in 1D; therefore no filter is applied here. The decomposition of the perturbations into different wave modes becomes more complicated in 2D, and the details are described in Appendix~\ref{appen C}. The measurement is done within $t = 1 \times 10^4 \Omega_c^{-1}$. 

The measured growth rates of Alfven waves, fast waves and slow waves are shown in Figure~\ref{fig:2d_alfven}, Figure~\ref{fig:2d_fast} and Figure~\ref{fig:2d_slow}. Due to the system’s rotational symmetry about $\hat{x}$, waves in the first $(|k_x|, |k_y|)$ and fourth $(|k_x|, -|k_y|)$ quadrants of the $\boldsymbol{k}$-space have identical growth rates, so we show results only for the first quadrant. Waves in the second and third quadrants are backward-propagating and damped, thus are not shown here.

At all values of $\beta$, the growth rates measured in the simulations match the theory quite well over almost the full $k_x \text{--} k_y$ plane, as seen more quantitatively from the right panels, which show growth rates along selected slices. 
In Figure~\ref{fig:2d_alfven}, the measured growth rates in three 2D panels are similar, supporting our previous claim that the growth rates of Alfven waves are independent of $\beta$. Despite the similarity, the noise levels of highly oblique ($\theta \geq \pi/4$) waves increase with increasing $\beta$. A similar phenomenon is observed in Figure~\ref{fig:1d_sim} as well. 
In Figure~\ref{fig:2d_fast}, one particularly important finding is that our simulations accurately capture the critical angle phenomenon of fast waves. Besides, both the growth rate spectra and the critical angles clearly exhibit a dependence on $\beta$. The contours of theoretical growth rates suggest that, for certain values of $k_x$, oblique magnetosonic waves may have higher growth rates than parallel ones. Such behavior is indeed observed in our simulations, particularly in Figure~\ref{fig:2d_slow}. Similar to the 1D simulations, in 2D simulations, the measured growth rates are truncated around $k_x \sim 10 m\Omega_c / p_0$, which is also due to the numerical dissipation. 
While we did not do 3D simulations, the physics of oblique waves remains the same in 3D, and we expect a similar agreement between the theory and the simulations.

\subsubsection{Quasi-linear Evolution and the Saturated State}
\begin{figure*}
    \includegraphics[width=17cm]{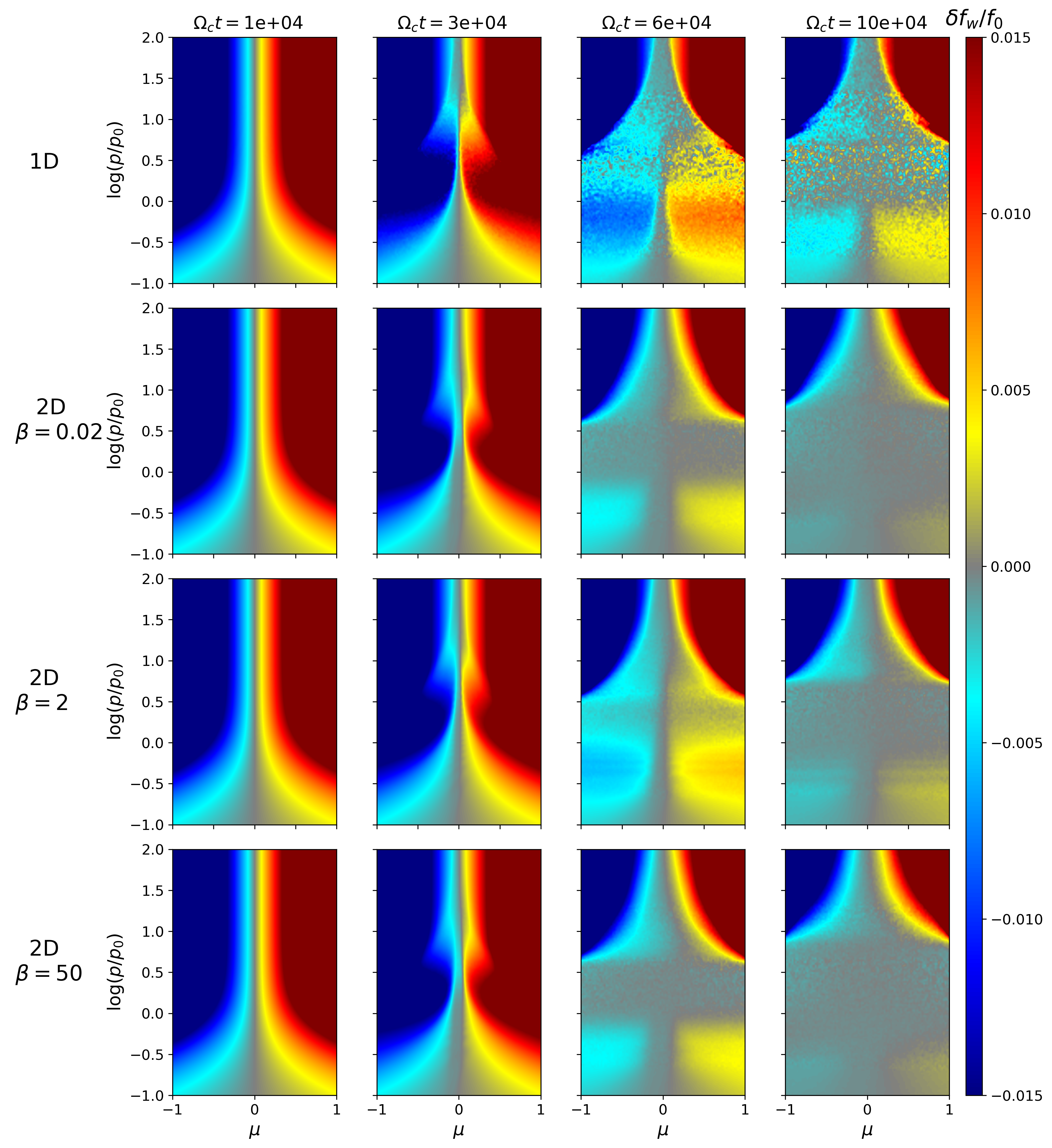}
    \caption{The 2D distribution function $\delta f_w / f_0$ in the wave frame at four snapshots in run 1D-c and 2D runs. Here $\mu$ is the pitch angle cosine.}
    \label{fig:evo_dis}
\end{figure*}
\begin{figure}
    \includegraphics[width=\columnwidth]{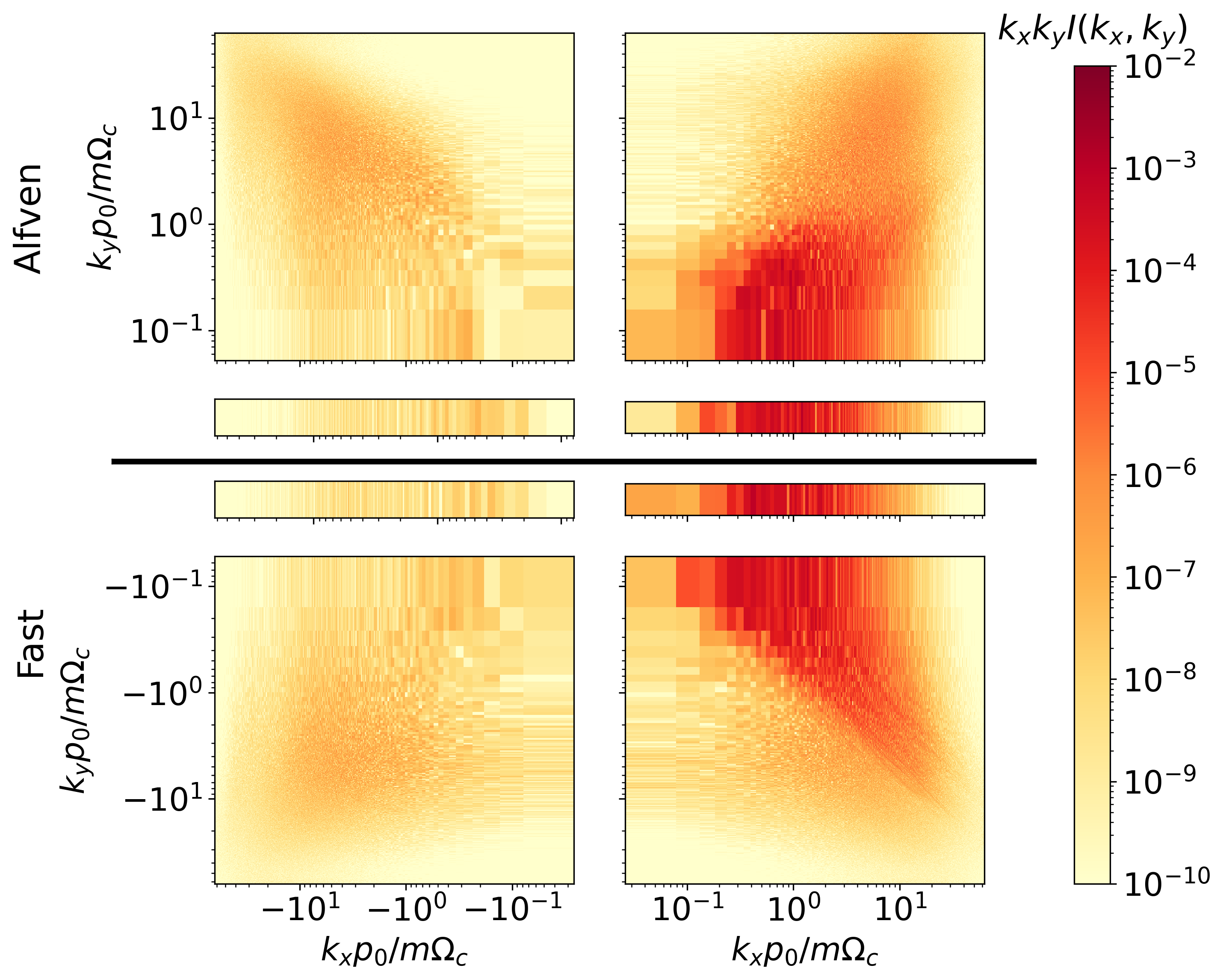}
    \caption{The saturated spectra of Alfven and fast waves measured in the 2D simulation with $\beta = 0.02$. Here we use a logarithmic scale for the $y$ axis, which can not extend to $k_y = 0$. Thus the growth rates of parallel waves ($k_y = 0$) is shown in the middle stripes. The spectra are symmetric about $k_x$, thus we only show the first and second quadrants for Alfven waves and the third and fourth quadrants for fast waves.}
    \label{fig:saturated_spec}
\end{figure}
After the linear growth phase, the energy in waves becomes significant enough so that waves can substantially influence the momentum distribution of particles, driving it towards isotropy in the wave frame through QLD. This process is shown in Figure~\ref{fig:evo_dis}, where we display the time evolution of the particle distribution function in the wave frame $(\delta f_w/f_0\equiv (f_w - f_0)/f_0)$ via four successive snapshots of each run. The distribution in the wave frame becomes isotropic when $\delta f_w$ approaches 0. In all cases, isotropization is first achieved for particles with $p \gtrsim p_0$, which resonate with the fastest-growing mode. At the end of our simulations at $t = 10^5\Omega_c^{-1}$, substantial isotropization is achieved for all particles with $p \lesssim 5p_0$, whereas for particles with $p >  5p_0$, only those with small pitch angles are isotropized. This is because the low intensity at long wavelengths and the finite size of the simulation box, which corresponds to 20 times the most unstable wavelength. At the late stage ($t \geq 6 \times 10^4 \Omega_c^{-1}$), isotropization of low-energy particles $(p<p_0)$ is more effective in 2D simulations, leading to higher saturated wave energy densities in Figure \ref{fig:evo_e}, particularly in the $\beta=0.02$ and $\beta = 50$ cases. As discussed in \cite{Bai2019}, the main obstacle from isotropization is to reflect particles across 90-degree pitch angle where QLD fails. In the next subsection, we will show that in 2D simulations, particles can cross the 90-degree barrier more efficiently thanks to the oblique waves.

In the absence of a driving source in our simulations, the system reaches saturation once isotropization of the distribution in the wave frame is achieved. The saturated wave intensity spectra in all 2D runs exhibit similar features, except that the types of waves that appear are different. As a representative example, Figure~\ref{fig:saturated_spec} shows the saturated spectra of Alfven and fast waves for the $\beta = 0.02$ case. As expected, forward-propagating waves grow significantly, with a notable region in $k$-space showing a substantial intensity. By comparing Figures~\ref{fig:2d_alfven} and \ref{fig:2d_fast} with Figures~\ref{fig:saturated_spec}, one can see that the saturated intensity correlates with the linear growth rates. However, it does not necessarily coincide with the linear growth rate as the final wave amplitudes depend on the interplay with QLD. Despite remaining at low amplitudes, waves with $k_x < 0$ emerge. Note that this is also found in 1D simulations \citep{Bai2019}, likely attributed to the development of rotational discontinuities \citep{Plotnikov_2021}.

\subsubsection{Overcoming the 90-degree Barrier}
\label{sec:90-degree barrier}
It is well-known that the QLD fails to account for particles' pitch angles crossing 90-degree, and several mechanisms have been proposed to reflect particles. Within QLT, by relaxing the magnetostatic approximation ($\omega/k \sim 0$), \cite{Schlickeiser1989} found that diffusion across 90-degree pitch angle can occur if both forward- and backward- propagating waves are present. Beyond QLT, efforts to alleviate this problem in general resort to either nonresonant scattering effects, e.g., magnetic mirroring \citep{Felice_2001}, or nonlinear effects which lead to resonance broadening \citep{Dupree1966, volk1973, Achterberg1981}. In recent numerical simulations, the main mechanisms are identified as mirroring reflection or reflection across the rotational discontinuities \citep{Holcomb_2019, Bai2019, Bambic_2021}. In this study, we find that low-energy ($p<p_0$) particles mainly undergo mirror-like reflections, which conserves magnetic moment $M \equiv p_\perp^2 / 2B$, while high-energy ($p>3p_0$) particles usually cross 90-degree pitch angle within a few gyroperiods experiencing abrupt changes in magnetic field.

To reveal the mechanism behind particle crossing of the 90-degree pitch angle, we trace the trajectories of a subsample of particles for a time interval of $100 \Omega_c^{-1}$ in the saturated state, during which no phase randomization is applied. We count the number of particles that have undergone reflection across 90-degree over this interval. Here we consider the particle pitch angle in the wave frame that is relative to the full magnetic field at the guiding centers. The pitch angle cosine in the wave frame is denoted as $\mu_w$. The magnetic field at the guiding center is obtained by applying a moving average filter to the magnetic field sequence experienced by a particle, with a window width of the gyroperiod. For simplicity, we adopt the non-relativistic gyroperiod, which is identical for all particles. To be counted, a particle must exhibit a sign change in $\mu_w$ across the interval, with $|\mu_w|$ exceeding $v_A/\mathbb{C} = 0.0033$ both before and after the crossing.

\begin{figure*}
    \includegraphics[width=17cm]{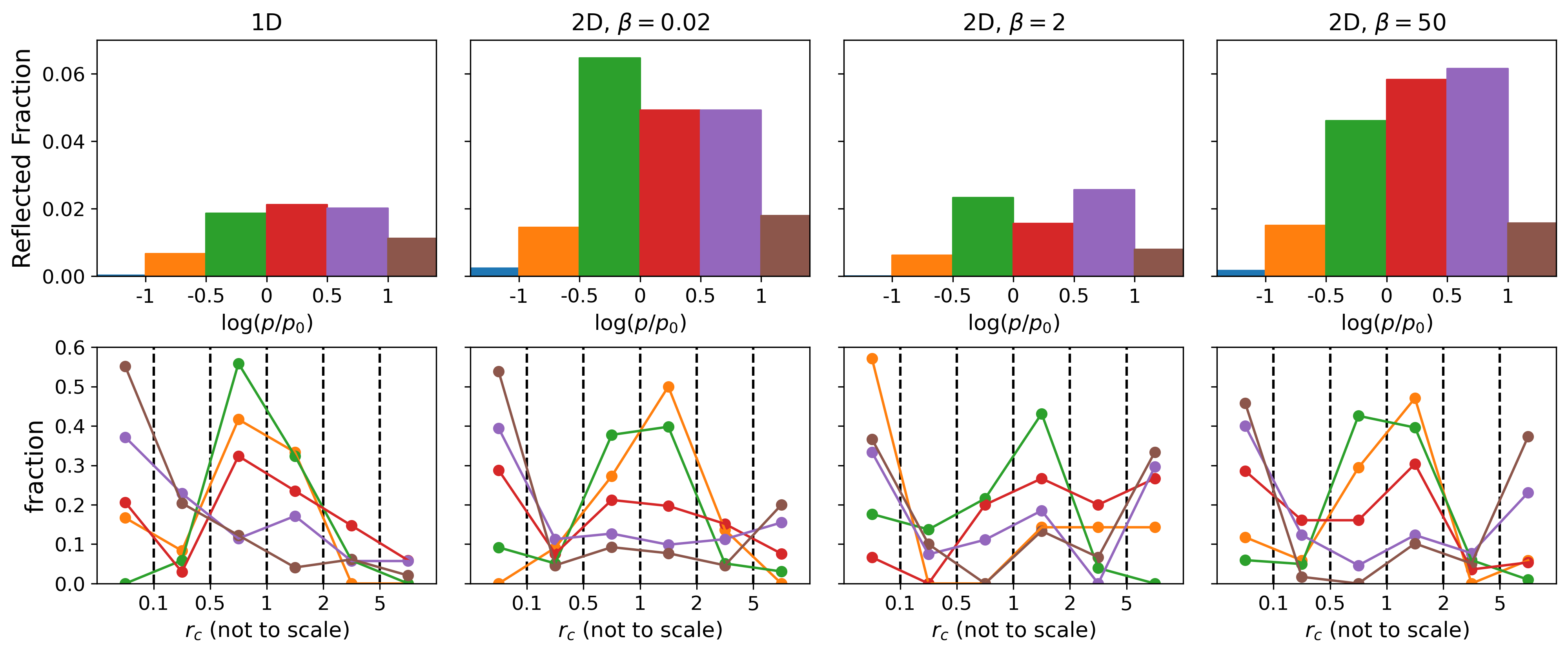}
    \caption{The fraction of particles that cross the 90-degree barrier (over a time inverval of 100$\Omega_c^{-1}$) in different momentum ranges (upper panels) and the distribution of $r_c$ for all reflection events within each momentum bin (bottom panel). In the bottom panels, the bins of $r_c$ are marked by vertical dashed black lines.}
    \label{fig:reflection_stat}
\end{figure*}
The fraction of particles that cross the 90-degree barrier in different momentum ranges is summarized in Figure~\ref{fig:reflection_stat}. In the 2D runs with $\beta = 0.02$ and $\beta = 50$, the crossing efficiency is comparable and approximately 2.4 times higher than in run 1D-c and the 2D run with $\beta = 2$, consistent with the enhanced isotropization observed in the former cases. Particles with momentum between $p_0/\sqrt{10}$ and $10 p_0$ exhibit a similar crossing efficiency, whereas those with higher or lower momenta show lower efficiency, likely due to the limited wave energy at high-$k$ and low-$k$.

\begin{figure}
    \includegraphics[width=\columnwidth]{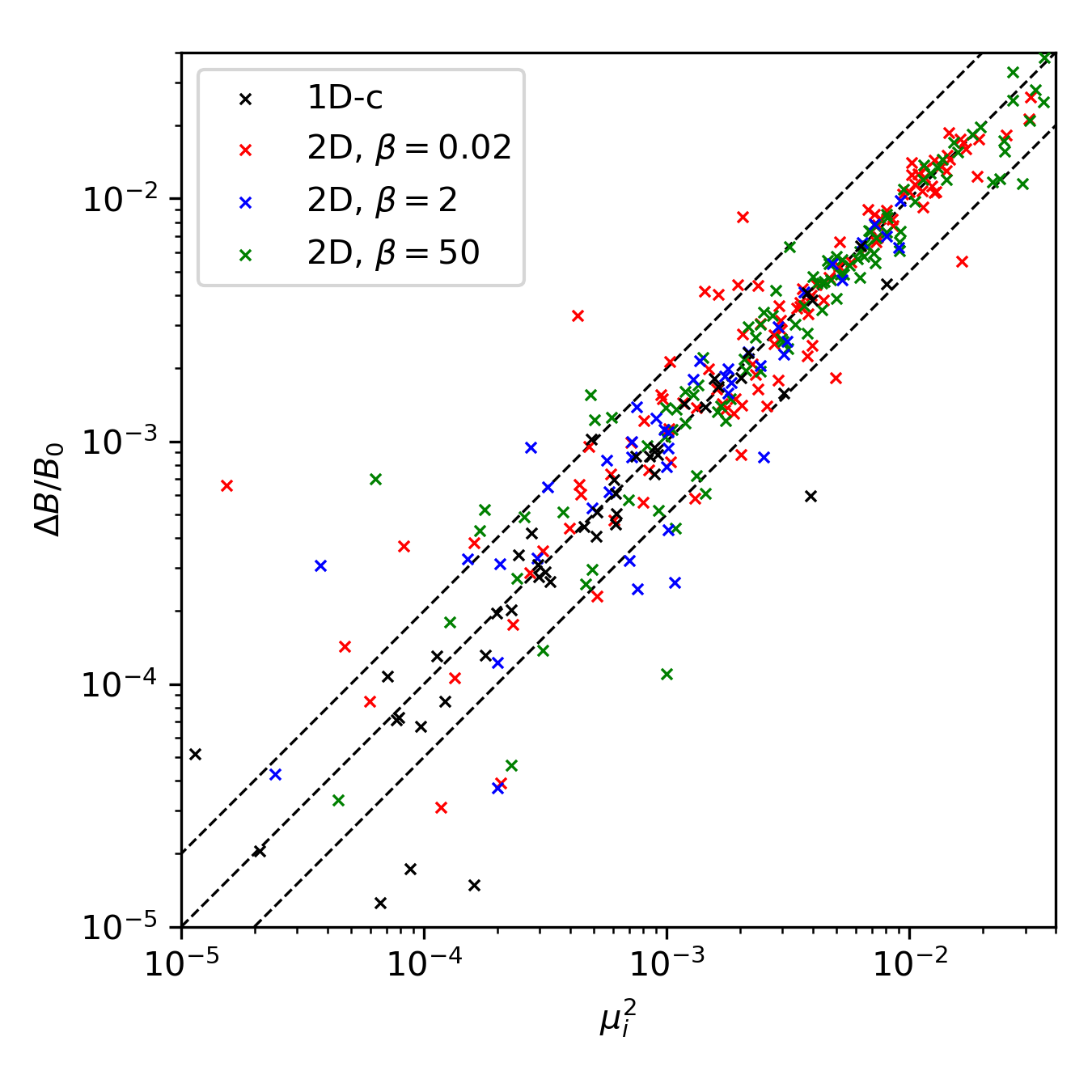}
    \caption{The correlation between $\mu_i^2$ and $\Delta B$ for low-energy particles ($0.1p_0 < p < p_0$) in run 1D-c and all 2D simulations.}
    \label{fig:mu_DeltaB}
\end{figure}
The mirroring and rotational discontinuity mechanisms can be distinguished by checking the behavior of magnetic moment during the reflection process. Initially, a particle has pitch angle cosine $\mu_i$ and experiences a magnetic field of strength $B_i$. Upon reflection, its pitch angle cosine becomes $\mu_f = 0$, and the magnetic field strength changes to $B_f$. We define the parameter $r_c \equiv (\Delta B / B_0) / \mu_i^2$, where $\Delta B \equiv B_f - B_i$. Assuming $\Delta B \ll B_f \approx B_0$, the conservation of the magnetic moment implies $r_c \approx 1$.
Figure~\ref{fig:reflection_stat} shows the distribution of $r_c$ for all reflection events across different momentum bins. In all runs, the distribution of $r_c$ shows a similar dependence on particle energy: For low-energy particles with momenta below $p_0$, the distribution peaks at $0.5 < r_c < 2$, indicating the conservation of magnetic moment during reflection, consistent with mirror-like reflections. For these particles, we further show the correlation between $\mu_i^2$ and $\Delta B/B_0$ in Figure~\ref{fig:mu_DeltaB}. Most data points lie along the $\mu_i^2 = \Delta B/B_0$ line, again supporting the mirror-like reflection. Besides, events with $\mu_i^2 \approx \Delta B/B_0 > 0.01$ are all from 2D runs with $\beta = 0.02$ and $\beta = 50$, indicating that larger magnetic field strength perturbations in these two runs allow particles to undergo reflection with higher $|\mu_i|$. As discussed at the beginning of this section, the field strength perturbation is primarily driven by compressible oblique magnetosonic waves. 
For higher-energy particles with momenta exceeding $p_0$, the distribution of $r_c$ is more diverse. The significant deviation of $r_c$ from unity for a substantial fraction of these particles indicates that mirror-like reflection is not the dominant reflection mechanism at higher energies.

\begin{figure*}
    \includegraphics[width=17cm]{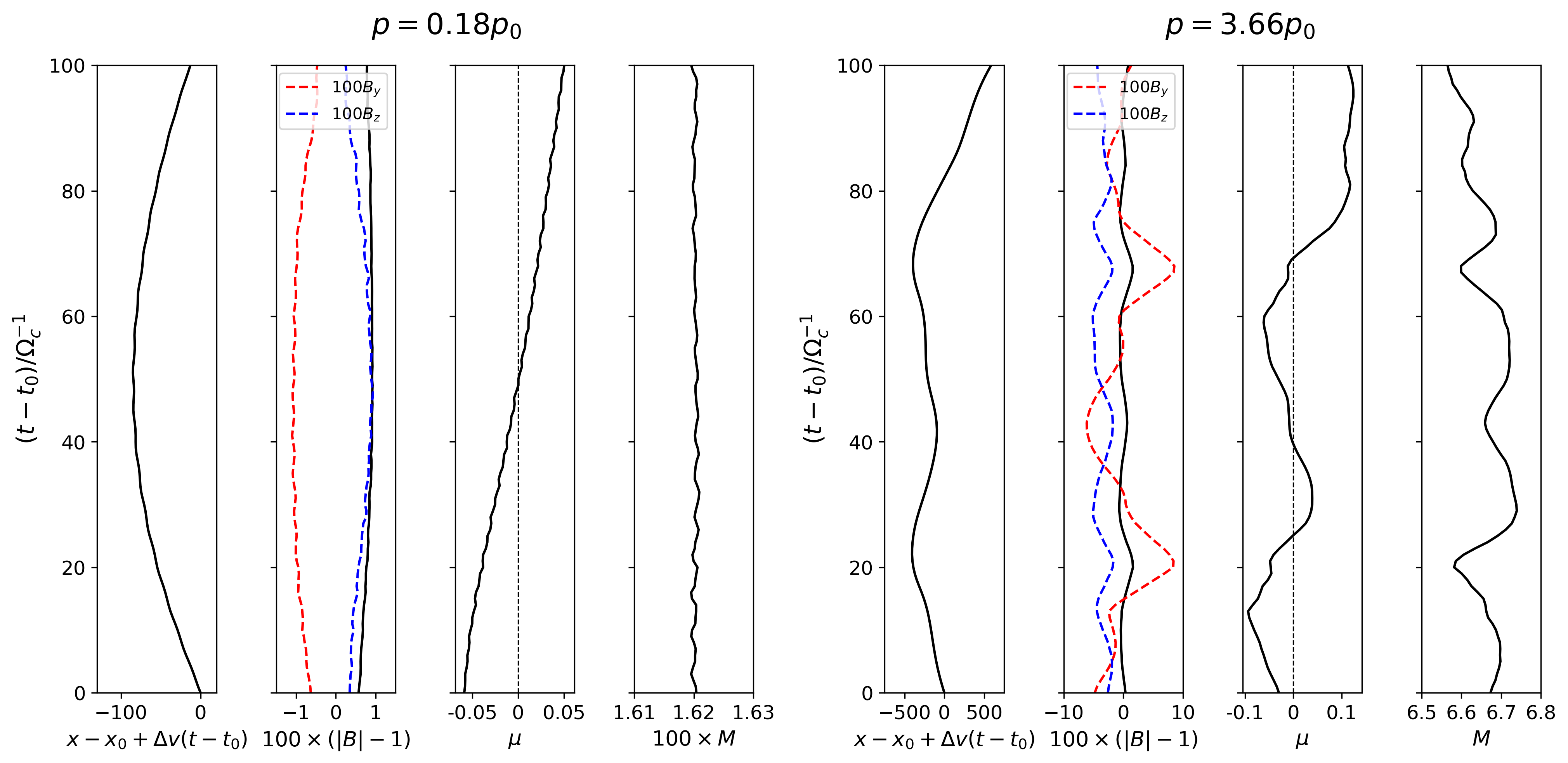}
    \caption{Two typical reflection events for particles with $p = 0.18 p_0$ (left) and $p = 3.66p_0$ (right). For each event, the four panels show the time evolution of the particle position, the field strength together with $B_y$ and $B_z$ experienced by the particle, its pitch angle cosine, and its magnetic moment.}
    \label{fig:traj}
\end{figure*}
To further confirm that low-energy particles undergo mirror-like reflections and investigate the mechanism that reflects higher-energy particles, we looked at the trajectories of reflected particles in 2D simulations and show two representative cases in Figure~\ref{fig:traj}. We find that most of the trajectories of low-energy particles exhibit the same characteristics: the pitch angle cosine changes smoothly and crosses zero when the magnetic field strength reaches maximum. Besides, the magnetic moment remains almost constant throughout the process. These features, illustrated in the left panel of Figure~\ref{fig:traj}, provide strong evidence that these particles indeed undergo mirror-like reflections. For higher-energy particles, we observe some particle reflection show mirror-like features, but not for other cases.\footnote{For these particles, $r_c$ becomes a less effective metric for determining whether or not they undergo mirror-like reflections, because previously it was defined based on $B_i$, $\mu_i$ at the initial (fixed) time step, not the actual time step when mirror-like mechanism starts to dominate.}
When not reflected by mirrors, we commonly see that the change in pitch angle cosine is more abrupt and accompanied by a significant variation in magnetic moment, as shown in the right panel of Figure~\ref{fig:traj}. Additionally, there are abrupt changes in $B_y$ and $B_z$. This is consistent with the mechanism identified in \cite{Bai2019}, which can be attributed to the formation of rotational discontinuities \citep{Plotnikov_2021, Bambic_2021}. As shown at the beginning of this section, weak oblique shocks form in 2D simulations, which could be counterparts of the rotational discontinuities. For low-energy particles, these shock structures enhance mirror-like reflections. We notice that there was a miscalculation of $\delta B$ in \cite{Bai2019}, who thus did not attribute the reflection of low-energy particles to mirrors. While the role of mirrors was studied for a single case in \cite{Holcomb_2019}, here we provide a more in-depth, energy-dependent analysis.

\section{Summary and Discussion}
\label{sec:dis and conc}
In this paper, we present the first systematic study of the CRSI in multi-dimensions, using both analytical and numerical approaches. We first present the dispersion relation of the CRSI for all MHD waves with arbitrary propagation directions under a general plasma $\beta$. This allows us to investigate the dependence of the growth rates on $\beta$ and the wave propagation direction $\theta$, with the following major findings:
\begin{enumerate}
    \item Waves with $\theta$ up to $\sim \pi/6$ exhibit growth rates comparable to those of parallel modes across the entire spectrum. For the fast wave, there is no growing mode when $\theta$ exceeds the critical angle $\theta_c$, which is determined by Equation~\ref{eq:theta_c},
    \begin{equation}
        \cos^2 \theta_c = (\frac{v_A}{v_d})^2 \, \bigg[ 1 + \frac{\beta}{2}\big(1 - (\frac{v_A}{v_d})^2\big)\bigg]. \notag
    \end{equation}
    \item The growth rate of Alfven waves is not affected by $\beta$; the growth rate of slow waves increases with increasing $\beta$; the growth rate of fast waves decreases with increasing $\beta$ and transitions to decay when $\beta$ exceeds the critical $\beta$, $\beta_c=2(v_d/v_A)^2$.
    \item For large values of $\theta$ $(\theta \gtrsim \pi/3)$, zeroth-order resonance dominates the growth of magnetosonic waves, leading to a growth rate significantly higher than that of Alfven waves. The condition of zeroth-order resonance, $\mu_w = 0$, indicates that it is closely related to the 90-degree barrier problem.
\end{enumerate}

We employ 1D and 2D MHD-PIC simulations to verify the analytical dispersion relations. These simulations cover a wide range of $\beta$ $(0.02\text{--}50)$ and $\theta$ (for 1D, $0.1\text{--}1.2$; for 2D, $0\text{--} \pi/2$). In all cases, we measure the growth rate on a $k$-by-$k$ basis, and find that our simulations well reproduce the analytical expectations.

We further study the quasi-linear evolution of the CRSI, during which the distribution function in the wave frame is gradually isotropized through QLD. In the saturated stage of our 2D simulations, the field lines wiggle but remain nearly parallel, with no obvious sign of magnetic mirror/bottle configuration. Oblique Alfven and magnetosonic waves propagate across the entire domain, interfere with each other, and form weak oblique shocks.
A particularly important finding is the better isotropization of low-energy $(p<p_0)$ particles in 2D simulations. By tracing a subsample of particles, we find that these low-energy particles mainly undergo mirror-like reflection, and the stronger field strength perturbations generated by oblique waves enable more efficient reflection. For higher energy $(p>p_0)$ particles, they experience abrupt changes in $B_y$ and $B_z$ upon reflection, which is likely related to the rotational discontinuities.

The saturated wave energy is lower at $\beta=2$ compared to $\beta=0.02$ and $\beta=50$, consistent with the lower isotropization level and lower efficiency of crossing the 90-degree barrier. In Figure~\ref{fig:growth_rate_gallery}, we note that although both fast and slow waves can grow at $\beta = 2$, their growth is slower than the dominant magnetosonic mode at other $\beta s$. It is natural to expect that this slower growth may lead to lower saturated amplitudes. Since magnetosonic waves play an important role in helping reflect particles, their lower saturated amplitudes reduce the crossing efficiency. However, we found that at $\beta = 2$, the Alfven wave energy and the magnetosonic wave energy start to fall below those at other $\beta$ values at the same time, $t \approx 3500 \Omega_c^{-1}$. This suggests that the lower energy level at $\beta = 2$ cannot be fully explained by linear theory, where the growth of Alfven waves should not be affected by $\beta$, though a more thorough investigation is beyond the scope of this work.

As a first step in the study of multidimensional effects in the CRSI, the theoretical derivations and simulations in this paper are limited to the framework of ideal MHD, without considering any wave-damping effects, which can be kinetic or fluid in nature, depending on the environment. These include ion-neutral damping \citep{Kulsrud1969}, nonlinear Landau damping \citep{Lee&Volk1973}, turbulent damping \citep{Farmer_2004, Lazarian_2016}, TTD for oblique waves \citep{Foote&Kulsrud1979}, etc. It can be expected that incorporating these damping mechanisms into the analysis would yield more realistic results. Additionally, all simulations in this paper employ periodic boundary conditions, which can be extended in future work with more realistic ones, such as the ``streaming box" framework employed in \cite{Bai_2022} to drive CR streaming under an imposed CR pressure gradient. This would allow us to offer a precise calibration of the CR scattering rates in the self-confinement regime. 

Among all the damping mechanisms mentioned above, TTD is of particular interest, since it is a linear mechanism and can be very strong, especially in high-$\beta$ plasmas. TTD is the magnetic analogue of Landau damping in which fluctuations in magnetic field strength caused by oblique magnetosonic waves resonantly interact with magnetic-moment-conserving particles. In low-$\beta$ plasmas, Alfven waves do not generate longitudinal electric or magnetic field perturbation, and thus are not subject to Landau damping or TTD. Assuming that electrons and ions are in thermal equilibrium, it is straightforward to show that electrons make the main contribution to the TTD of fast waves due to their higher thermal velocities. For simplicity, let us assume all ions are protons, then the damping rate is given by Equation~\ref{eq:growth rate},
\begin{align}
    \Gamma^{\mathrm{TTD}} &= -\frac{1}{2}\cos^2\alpha (\frac{v_A}{c})^2 \chi_{yy, I}^\mathrm{TTD} \, \omega \notag \\ &= -\frac{\sqrt{\pi}}{2\sqrt{2}}  \sqrt{\beta}\frac{k_\perp^2 v_A}{k_\parallel} \sqrt{\frac{m_e}{m_p}}\exp \bigg[-\frac{2m_e}{m_p} \frac{1}{\beta \cos^2\theta}\bigg].
    \label{eq:low_beta_TTD}
\end{align}
In the above, we have used $\cos^2\alpha \approx 1$ and $\omega_f = kv_A$ in the $\beta \ll 1$ limit. The contribution of TTD to the imaginary part of $\chi_{yy}$, $\chi_{yy, I}^{\mathrm{TTD}}$, has been calculated in \cite{Stix1992} for Maxwellian distributions. According to Equation~\ref{eq:grow rate final}, the growth rate of CRSI of fast waves is 
\begin{equation}
    \Gamma_f^{\mathrm{CRSI}} \approx (\frac{v_d \cos\theta}{v_A} - 1) \frac{n_\mathrm{CR}}{n_i} \Omega_c.
\end{equation}
This leads to
\begin{align}
    \frac{|\Gamma^{\mathrm{TTD}}|}{\Gamma_f^{\mathrm{CRSI}}} \approx 300& \bigg(\frac{v_A/c}{2\times10^{-5}}\bigg) 
    \bigg(\frac{n_i/n_{\mathrm{CR}}}{10^9}\bigg) \notag \\ 
    &\cdot \frac{\tan\theta \sin\theta}{(v_d\cos\theta/v_A - 1)}\sqrt{\beta} \exp \bigg[- \frac{10^{-3}}{\beta \cos^2\theta}\bigg],
\end{align}
where we have used $kv_A/\Omega_c \sim v_A/(r_L \Omega_c) \sim v_A/c$. For the typical values in the ISM, i.e., $B\sim3\mathrm{\mu G}$ and $n_i \sim 1 \mathrm{cm}^{-3}$, $v_A/c \sim 2\times 10^{-5}$. Assuming that $(v_d \cos\theta /v_A - 1) \sim 1$ and $n_i/n_{\mathrm{CR}} \sim 10^9$, for $\beta=0.01$, $|\Gamma^{\mathrm{TTD}}|/\Gamma_f^{\mathrm{CRSI}} \leq 1$ yields $\theta \leq 0.2 \, \mathrm{rad}=11\, \mathrm{deg}$; for $\theta=\pi/4$, $|\Gamma^{\mathrm{TTD}}|/\Gamma_f^{\mathrm{CRSI}} \leq 1$ yields $\beta \leq 1\times10^{-3}$. 

In high-$\beta$ plasmas, both Alfven waves and magnetosonic waves (slow waves in our context) are damped, and their damping rates have been calculated in detail in \citep{Foote&Kulsrud1979}. Unlike in low-$\beta$ plasma, where the eigenvector of the Alfven mode is simply $E_x$ and that of the fast mode is simply $E_y$, the eigenvectors of the Alfven and slow modes in high-$\beta$ plasmas become a mixture of $E_x$ and $E_y$. This mixing arises because the off-diagonal term of the susceptibility tensor, $\chi_{xy}$, becomes non-negligible due to the thermal motion of the ions. Consequently, since the eigenvector of the Alfven mode has a nonzero $E_y$ component, it undergoes damping. Since there are many thermal particles moving at the wave phase speed in a hot plasma, the exponential term in Equation~\ref{eq:low_beta_TTD} no longer exists, and the damping is generally strong. For $\theta \ll 1$ and $kr_i<v_A/v_i$, the damping rate of both modes can be approximated as
\begin{equation}
    \Gamma^{\mathrm{TTD}} \approx -\frac{\sqrt{\pi}}{4}k v_i \tan^2\theta,
\end{equation}
where $v_i$ is the ion thermal velocity and $r_i$ is the thermal ion gyroradius \citep{Foote&Kulsrud1979, Zweibel2017}. In the $\beta \gg 1$ limit, the growth rate of CRSI of both Alfven and slow waves may be approximated as $\Gamma_{a, s}^{\mathrm{CRSI}} \approx (v_d/v_A - 1) (n_\mathrm{CR}/n_i) \Omega_c$, and 
\begin{equation}
    \frac{|\Gamma^{\mathrm{TTD}}|}{\Gamma^{\mathrm{CRSI}}_{a,s}} \approx \frac{\sqrt{\pi}}{4} \frac{1}{v_d/v_A - 1}\frac{v_i}{c}\frac{n_i}{n_\mathrm{CR}} \tan^2\theta,
\end{equation}
which is beyond order unity except for near-parallel waves under typical conditions. 

Overall, we see that in low-$\beta$ plasmas, oblique Alfven waves are not subject to Landau damping or TTD, whereas oblique fast waves can grow only in plasmas with $\beta \lesssim 10^{-3}$ under typical conditions. In high-$\beta$ plasmas, nearly all oblique waves are strongly damped unless under exceptional conditions (e.g., very high $n_{\rm CR}/n_i$). Therefore, oblique waves are in general expected to be important primarily in low-$\beta$ plasmas.

Finally, we note that multidimensional simulations in principle enable the measurement of perpendicular diffusion from the CRSI. We attempted to do so
by turning off phase randomization and tracking the perpendicular displacements $(\delta y)$ of a subsample of particles from $t = 8\times 10^4\Omega_c^{-1}$ to $t = 10 \times 10^4 \Omega_c^{-1}$. For CRs with momenta around $p_0$, we found that the mean square displacement, $\langle(\delta y)^2\rangle$, grew in the first $20\Omega_c^{-1}$, and then remained constant. The running perpendicular diffusion coefficient at the end of our tracking is
\begin{equation}
    D_{yy}(\Delta t=2\times 10^4\Omega_c^{-1}) = \frac{\langle(\delta y)^2\rangle}{2\Delta t} \leq 4.4\times 10^{-7} r_L^2 \Omega_c,
\end{equation}
where $r_L = p_0/m\Omega_c = 300 d_i$, and the $"\leq"$ comes from the uncertainty of the origin of perpendicular displacements. In comparison, the parallel diffusion coefficient may be estimated as
\begin{equation}
    D_{xx} \approx \frac{1}{3} \bigg(\frac{B_0}{\delta B} \bigg)^2 r_L^2 \Omega_c \approx 3.3 \times 10^2 r_L^2 \Omega_c.
\end{equation}
This suggests that perpendicular diffusion is weak in the quasi-linear regime, but it is also likely that our current setup, particularly the limited box size with periodic boundary condition, is not well-suited for studying perpendicular diffusion. Thus, we leave the investigation of perpendicular diffusion for future work.

\begin{acknowledgments}
We thank the referee and Brian Reville for helpful comments and suggestions on the manuscript. S.Z. thanks Xinle Cheng and Xihui Zhao for insightful discussions. This work is supported by National Science Foundation of China under grant No. 12325304, 12342501. X. S. acknowledges the support from Multimessenger Plasma Physics Center (MPPC, NSF grant PHY-2206607). Numerical simulations are conducted in the Orion cluster at Department of Astronomy, Tsinghua University, and the Zaratan cluster which is supported by the Division of Information Technology at the University of Maryland.
\end{acknowledgments}

\appendix

\section{General Analysis of the Eigen Modes}
\label{appen A}
The full linearized CR modified MHD equations read,
\begin{align}
    \frac{\partial \rho}{\partial t} + &\rho_0 \nabla \cdot \boldsymbol{u} = 0  \\
    \rho_0 \frac{\partial \boldsymbol{u}}{\partial t} &= \frac{1}{4 \pi}(\nabla \times \boldsymbol{B})\times \boldsymbol{B_0} - \frac{1}{c} \boldsymbol{j}_{\mathrm{CR},0} \times \boldsymbol{B} \notag \\ & - \frac{1}{c} \boldsymbol{j}_{\mathrm{CR},1} \times \boldsymbol{B_0} - e n_{\mathrm{CR}} \boldsymbol{E} - \nabla P  \\
    \frac{\partial \boldsymbol{B}}{\partial t} &= \nabla \times (\boldsymbol{u} \times \boldsymbol{B_0})  \\
    \delta P &= c_s^2 \delta \rho .
\end{align}
Again consider perturbations of the form $\exp(\mathrm{i} (\boldsymbol{k} \cdot \boldsymbol{r} - \omega t))$, then the mass, momentum and induction equations become 
\begin{align}
    - \mathrm{i} \omega \delta \rho &+ \rho_0 \mathrm{i} \boldsymbol{k} \cdot \boldsymbol{u} = 0 \\
    - \mathrm{i} \omega \boldsymbol{u} &= \frac{\mathrm{i}}{4 \pi \rho_0} (\boldsymbol{k} \times \delta\boldsymbol{B}) \times \boldsymbol{B}_0 - \frac{1}{\rho_0 c} \boldsymbol{j}_{\mathrm{CR},0} \times \delta\boldsymbol{B} \notag \\
    & - \frac{1}{\rho_0 c} \boldsymbol{j}_{\mathrm{CR},1} \times \boldsymbol{B}_0 - \frac{e n_{\mathrm{CR}}}{\rho_0} \boldsymbol{E} - \frac{\mathrm{i}}{\rho_0}\boldsymbol{k} \delta P, \\
    - \mathrm{i} \omega \delta\boldsymbol{B} &= \mathrm{i} \boldsymbol{k} \times (\boldsymbol{u} \times \boldsymbol{B_0}).
\end{align}
Using $\boldsymbol{E} = -\boldsymbol{u} \times \boldsymbol{B_0}/c$ and $\delta P = c_s^2 \delta \rho = c_s^2 \rho_0 \boldsymbol{k} \cdot \boldsymbol{u}/\omega$, the momentum equation can be rewritten as 
\begin{align}
    - \mathrm{i} \omega \boldsymbol{u} = &- \frac{\mathrm{i}}{4 \pi \rho_0} \left[\boldsymbol{k} \times (\frac{\boldsymbol{k}}{\omega} \times (\boldsymbol{u} \times \boldsymbol{B_0}))\right] \times \boldsymbol{B}_0 \notag \\
    &+ \frac{1}{\rho_0 c} \boldsymbol{j}_{\mathrm{CR},0} \times \frac{1}{\omega}\left[\boldsymbol{k} \times (\boldsymbol{u} \times \boldsymbol{B_0})\right] - \frac{e n_{\mathrm{CR}}}{\rho_0 c} (\boldsymbol{B_0} \times \boldsymbol{u}) \notag \\
    &- \frac{1}{\rho_0 c} \boldsymbol{B_0} \times \left[\frac{\mathrm{i} \omega}{4\pi} \boldsymbol{\chi}^{\mathrm{CR}} \cdot \frac{1}{c}(\boldsymbol{B_0} \times \boldsymbol{u})\right] - \frac{\mathrm{i} c_s^2}{\omega}\boldsymbol{k} (\boldsymbol{k} \cdot \boldsymbol{u}).
\end{align}
This is a linear equation of $\boldsymbol{u}$ and we rewrite it into the matrix form
\begin{equation}
    \boldsymbol{T} \cdot \boldsymbol{u} = 0.
\end{equation}
After some algebra, we can get the explicit form of $\boldsymbol{T}$:

\begin{align}
T_{xx} &= \frac{v_A^2 k^2}{\omega} - \frac{v_A^2 \omega}{c^2}\chi^{\mathrm{CR}}_{yy} + \frac{c_s^2 k_\perp^2}{\omega} - \omega, \\
T_{xy} &= -\mathrm{i} \frac{j_{\mathrm{CR}, 0} B_0 k_\parallel}{\rho_0 c \omega} + \frac{v_A^2 \omega}{c^2} \chi^{\mathrm{CR}}_{yx} + \frac{\mathrm{i} e n_{\mathrm{CR}} B_0}{\rho_0 c}, \\
T_{xz} &= T_{zx} = \frac{c_s^2}{\omega} k_\perp k_\parallel, \\
T_{yx} &= \mathrm{i} \frac{j_{\mathrm{CR}, 0} B_0 k_\parallel}{\rho_0 c \omega} + \frac{v_A^2 \omega}{c^2} \chi^{\mathrm{CR}}_{xy} - \frac{\mathrm{i} e n_{\mathrm{CR}} B_0}{\rho_0 c}, \\
T_{yy} &= \frac{v_A^2 k_\parallel^2}{\omega} - \frac{v_A^2 \omega}{c^2}\chi^{\mathrm{CR}}_{xx} - \omega, \\
T{yz} &= T_{zy} = 0, \\
T_{zz} &= \frac{c_s^2 k_\parallel^2}{\omega} -\omega.
\end{align}
The dispersion relation is given by $\det \boldsymbol{T} = 0$, and eigen modes are solutions of the degenerate equations.

Now let us consider the $n_{\mathrm{CR}}/n_i \ll 1$ limit and keep CR terms to the first order.
Expanding the determinant along the second row or column, it is straightforward to verify that the term associated with $T_{yx}$ is second order, while only the term associated with $T_{yy}$ is first order.
Therefore 
\begin{align}
    \det &\boldsymbol{T} = (\frac{v_A^2 k_\parallel^2}{\omega} - \frac{v_A^2 \omega}{c^2}\chi^{\mathrm{CR}}_{xx} - \omega) \notag \\
    \cdot & \bigg[(\frac{v_A^2 k^2}{\omega} - \frac{v_A^2 \omega}{c^2}\chi^{\mathrm{CR}}_{yy} + \frac{c_s^2 k_\perp^2}{\omega} - \omega)(\frac{c_s^2}{\omega}k_\parallel^2 - \omega) - \frac{c_s^4}{\omega^2} k_\perp^2 k_\parallel^2\bigg] 
    \label{eq:T}
\end{align}
Further assuming $\Gamma_i \ll \omega$ and expanding Equation ~\ref{eq:T} to the first order of $\Gamma$, we get
\begin{equation}
    \Gamma_i = -\frac{\omega}{2} (\frac{v_A}{c})^2 \left\{
    \begin{array}{c}
    \chi^{\mathrm{CR}}_{xx, I} \\[6pt]
    \dfrac{1}{2} \bigg(1 + \dfrac{1 - \beta \cos(2\theta)/2}{\sqrt{(1+\beta/2)^2 - 2\beta \cos^2 \theta}}\bigg) \, \chi^{\mathrm{CR}}_{yy, I} \\[6pt]
    \dfrac{1}{2} \bigg(1 - \dfrac{1 - \beta \cos(2\theta)/2}{\sqrt{(1+\beta/2)^2 - 2\beta \cos^2 \theta}}\bigg) \, \chi^{\mathrm{CR}}_{yy, I}
    \end{array} \right\}.
    \label{eq:growth rate again}
\end{equation}
It is straightforward to show that 
\begin{align}
    \cos^2\alpha &= \frac{1}{2} \bigg(1 + \frac{1 - \beta \cos(2\theta)/2}{\sqrt{(1+\beta/2)^2 - 2\beta \cos^2 \theta}}\bigg), \\
    \sin^2\alpha &= \frac{1}{2} \bigg(1 - \frac{1 - \beta \cos(2\theta)/2}{\sqrt{(1+\beta/2)^2 - 2\beta \cos^2 \theta}}\bigg),
\end{align}
and Equation~\ref{eq:growth rate again} and Equation~\ref{eq:growth rate} are identical.

\begin{figure}
    \includegraphics[width=\columnwidth]{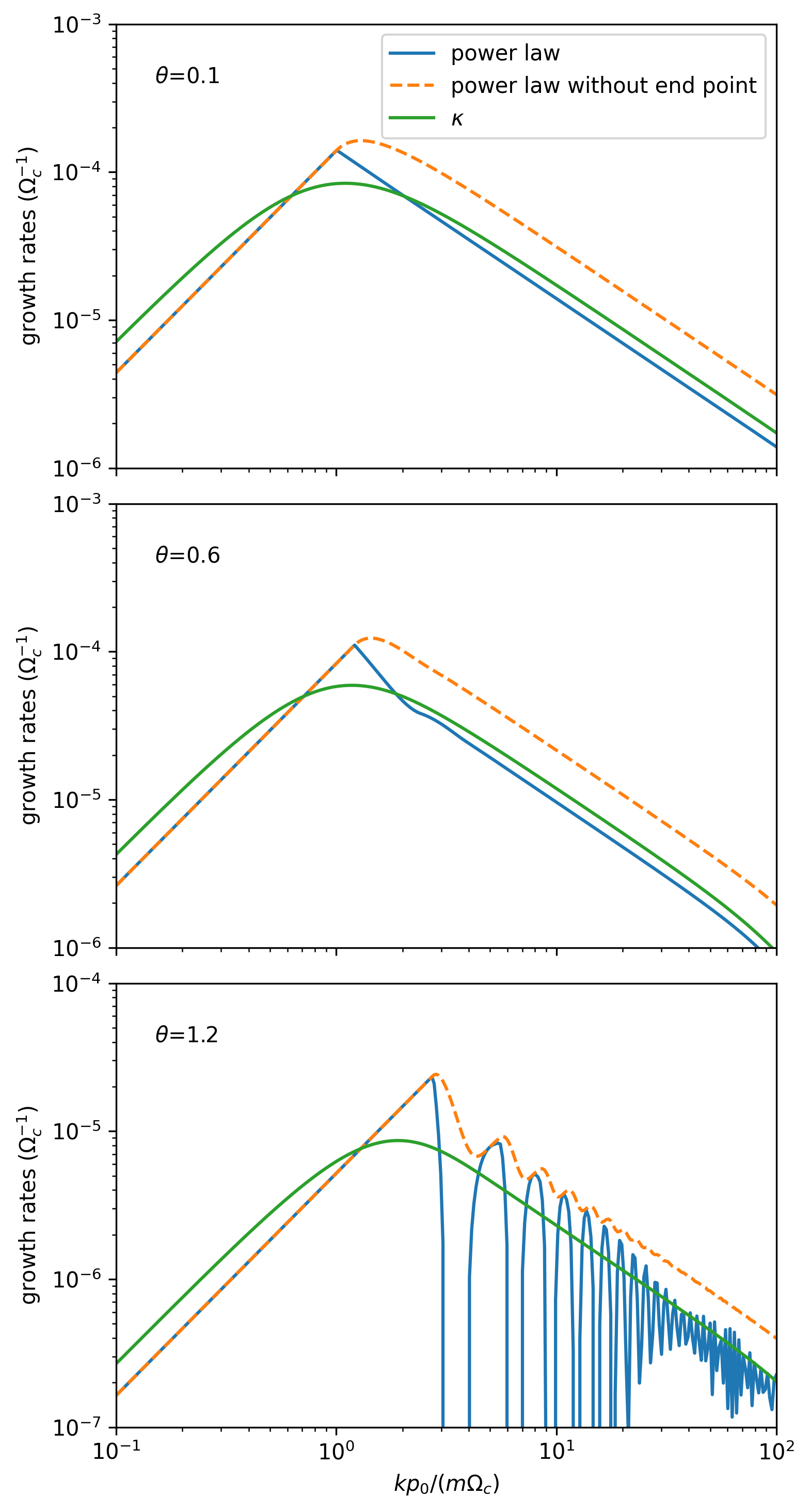}
    \caption{The growth rates of truncated power law distribution with and without the contribution of the discontinuity at the boundaries. For comparison, we also plot the growth rates of $\kappa$ distribution.}
    \label{fig:discontin}
\end{figure}
\section{Effect of the Discontinuity of Truncated Power Law Distribution}
\label{appen B}
In this appendix, we show that the artificial discontinuity of truncated power law distribution could lead to spurious growth rates when $\theta$ is large. For simplicity, we take Alfven mode as an example below.

The truncated power law distribution reads,
\begin{equation}
    F(p) \propto p^{-(4+\alpha)} \eta(p - p_0),
\end{equation}
where $\eta$ is the step function. Derivative of this distribution is 
\begin{equation}
    \frac{\mathrm{d}F}{\mathrm{d}p} \propto \left[-(4+\alpha)p^{-(5+\alpha)} \eta(p - p_0) + p^{-(4+\alpha)}\delta(p-p_0)\right].
\end{equation}
In the above equation, the appearance of the $\delta$-function is due to the discontinuity of the distribution function, and we could investigate its effect by comparing the growth rates including it or not. The results are shown in Figure~\ref{fig:discontin}, where we also plot the growth rates of $\kappa$ distribution for comparison.

From this figure, it is clear that when $\theta$ is large, the discontinuity leads to significant oscillations in growth rates in the high-$k$ regime. When excluding the $\delta$-function term, the resulting growth rates of the power law distribution are similar to those of the $\kappa$ distribution. Therefore, we consider the $\kappa$ distribution as a more reasonable and realistic one and adopt it in this work.

\section{Eigenvectors and Mode Decomposition}
\label{appen C}
In this appendix, we provide the explicit form of the eigenvalues and eigenvectors of our system, and the method of wave decomposition. The former part largely follows Appendix A of \cite{Stone_2008}. As a general discussion, we consider homogeneous gas with drift velocity $\boldsymbol{v_d} = (v_{x0}, v_{y0}, v_{z0}$) and magnetic field $\boldsymbol{B_0} = (B_{x0}, B_{y0}, B_{z0})$. For perturbations along $x$ axis, the linearized equation of motion reads
\begin{equation}
    \frac{\partial W_1}{\partial t} + \Lambda \frac{\partial W_1}{\partial x} = 0,
\end{equation}
where $W_1$ is a vector composed of the perturbed primitive variables (hereafter the subscript $1$ means perturbed fields) and $\Lambda$ is a matrix. The eigenvectors of $\Lambda$ are the wave modes of this system while the eigenvalues are the corresponding wave speeds. For isothermal MHD, $W_1 = (\rho_1, u_{x1}, u_{y1}, u_{z1}, b_{y1}\equiv B_{y1}/\sqrt{4\pi}, b_{z1}\equiv B_{z1}/\sqrt{4\pi})$,
\begin{equation}
    \boldsymbol{\Lambda} = 
        \begin{bmatrix}
            v_{x0} & \rho_0 & 0 & 0 & 0 & 0 \\
            c_s^2/\rho_0 & v_{x0} & 0 & 0 & b_{y0}/\rho_0 & b_{z0}/\rho_0 \\
            0 & 0 & v_{x0} & 0 & -b_{x0}/\rho_0 & 0 \\
            0 & 0 & 0 & v_{x0} & 0 & -b_{x0}/\rho_0 \\
            0 & b_{y0} & -b_{x0} & 0 & v_{x0} & 0 \\
            0 & b_{z0} & 0 & -b_{x0} & 0 & v_{x0}
        \end{bmatrix}.
\end{equation}
The six eigenvalues of this matrix in ascending order are 
\begin{equation}
    \lambda = (v_{x0} - v_f, v_{x0} - v_{Ax}, v_{x0} - v_s, v_{x0} + v_s, v_{x0} + v_{Ax}, v_{x0} + v_f),
\end{equation}
where 
\begin{equation}
    v_{f, s} = \frac{1}{2} [(c_s^2 + v_A^2 \pm \sqrt{(c_s^2 + v_A^2)^2 - 4 c_s^2 v_{Ax}^2}], v_{Ax}^2 = b_{x0}^2/\rho_0.
\end{equation}
The eigenvectors of $\Lambda$ are the columns of the following matrix,
\begin{equation}
    \boldsymbol{R} = 
        \begin{bmatrix}
            \rho_0 \chi_f & 0 & \rho_0 \chi_s & \rho_0 \chi_s & 0 & \rho_0 \chi_f \\
            -C_{ff} & 0 & -C_{ss} & C_{ss} & 0 & C_{ff} \\
            Q_s \beta_y & -\beta_z v_A & -Q_f\beta_y & Q_f\beta_y & \beta_z v_A & -Q_s\beta_y \\
            Q_s \beta_z & \beta_y v_A & -Q_f\beta_z & Q_f\beta_z & -\beta_y v_A & -Q_s\beta_z \\
            A_s\beta_y & -\beta_z S b_0 & -A_f \beta_y & -A_f \beta_y & -\beta_z S b_0 & A_s \beta_y \\
            A_s\beta_z & \beta_y S b_0 & -A_f \beta_z & -A_f \beta_z & \beta_y S b_0 & A_s \beta_z
        \end{bmatrix},
\end{equation}
where $S = \mathrm{sgn}(b_x)$, and 
\begin{align}
    C_{ff} = v_f \chi_f&, \quad C_{ss} = v_s \chi_s, \notag \\
    Q_f = v_f \chi_f S&, \quad Q_s = v_s \chi_s S, \notag \\
    A_f = c_s \chi_f \sqrt{\rho_0}&, \quad A_s = c_s \chi_s \sqrt{\rho_0}, \notag \\
    \chi_f^2 = \frac{c_s^2 - v_s^2}{v_f^2 - v_s^2} \frac{v_A^2}{c_s^2}&, \quad \chi_s^2 = \frac{v_f^2 - c_s^2}{v_f^2 - v_s^2} \frac{v_A^2}{c_s^2}, \notag \\
    \beta_y = \frac{b_{y0}}{\sqrt{b_{y0}^2 + b_{z0}^2}}&, \quad \beta_z = \frac{b_{z0}}{\sqrt{b_{y0}^2 + b_{z0}^2}}.
\end{align}
In the above, we normalize all eigenvectors such that $u_1^2 \equiv u_{x1}^2 + u_{y1}^2 + u_{z1}^2 = v_A^2$. With this choice, energy density of a single wave is simply $\epsilon = \rho_0 v_A^2 A^2/2$, where $A$ is its amplitude, and independent of the isothermal sound speed $c_s$ we choose in simulations.

In post-processing, we need to decompose the perturbation fields $W_1$ into a combination of the above eigen modes, i.e.
\begin{equation}
    W_1 = R a,
\end{equation}
where $a$ is the vector composed of the amplitudes of different wave modes. Given that the eigenvectors are independent, $R$ is reversible and the decomposition can be done by 
\begin{equation}
    a = R^{-1} W_1.
    \label{eq:general_decomp}
\end{equation}

In our simulations, by setting $\boldsymbol{B_0}$ in the $x\text{--} z$ plane, we have $\beta_y = 0$ and $\beta_z = 1$. We further reorganize $W_1$ as $W_1 = (W_a, W_{fs}) \equiv ((v_{y1}, b_{y1}), (\rho_1, v_{x1}, v_{y1}, b_{z1}))$, then $\boldsymbol{R}$ become a block diagonal matrix,
\begin{equation}
    \boldsymbol{R} = 
        \begin{bmatrix}
            \boldsymbol{R}_a & \boldsymbol{0} \\
            \boldsymbol{0} & \boldsymbol{R}_{ms} 
        \end{bmatrix},
\end{equation}
where
\begin{equation}
    \boldsymbol{R}_a = 
        \begin{bmatrix}
            -v_A & v_A \\
            -Sb_0  & -Sb_0 
        \end{bmatrix},
    \boldsymbol{R}_{ms} = 
        \begin{bmatrix}
            \rho_0 \chi_f & \rho_0 \chi_s & \rho_0 \chi_s & \rho_0 \chi_f \\
            -C_{ff} & -C_{ss} & C_{ss} & C_{ff} \\
            Q_s & -Q_f & Q_f & -Q_s \\
            A_s & -A_f & -A_f & A_s
        \end{bmatrix}.
\end{equation}
It is clear that $W_a$ and $\boldsymbol{R}_a$ correspond to the Alfven mode while $W_{ms}$ and $\boldsymbol{R}_{ms}$ correspond to the magnetosonic modes. In this way, Equation ~\ref{eq:general_decomp} can be simplified as 
\begin{equation}
    a_i = R_i^{-1} W_i,
\end{equation}
where $i$ can be $a$ or $ms$.

For 1D simulations, since the waves are set along $\hat{x}'$, the aforementioned method can be applied directly. For 2D simulations, since there are waves propagating along all directions, we first apply a Fourier transform to $W_1$, $\mathscr{W}_1(\boldsymbol{k}) = \mathscr{F}[W_1(\boldsymbol{r})]$. The resulting $\mathscr{W}_1(\boldsymbol{k})$ represents the perturbations propagating along $\boldsymbol{k}$, and thus can be decomposed into different wave modes using the above logic.

\nocite{*}

\bibliography{apssamp}

\end{document}